\documentclass[11pt]{article}
\usepackage{jheppub}

\makeatletter
\def\@fpheader{\relax}
\makeatother

\usepackage{dsfont}  % \mathds{1}
\usepackage{enumitem} % \begin{itemize}[leftmargin=*,noitemsep]

\usepackage{arydshln} % dashed lines in matrices  \hdashline
\usepackage{tocvsec2}
\settocdepth{subsection}
\title{Entanglement entropy of squeezed vacua on a lattice}

\author{Eugenio Bianchi,\;}
\emailAdd{ebianchi@gravity.psu.edu}
\author{Lucas Hackl,\;}
\emailAdd{lucas.hackl@psu.edu}
\author{Nelson Yokomizo}
\emailAdd{yokomizo@gravity.psu.edu}

\affiliation{Institute for Gravitation and the Cosmos \& Physics Department,\\ Penn State, University Park, PA 16802, USA}

\abstract{We derive a formula for the entanglement entropy of squeezed states on a lattice in terms of the complex structure $J$. The analysis involves the identification of squeezed states with group-theoretical coherent states of the symplectic group and the relation between the coset $Sp(2N,\mathbb{R})/\text{\emph{Isot}}(J_0)$ and the space of complex structures. We present two applications of the new formula: (i) we derive the area law for the ground state of a scalar field on a generic lattice in the limit of small speed of sound, (ii) we compute the rate of growth of the entanglement entropy in the presence of an instability and show that it is asymptotically bounded from above by the Kolmogorov-Sinai rate.}

\begin{document}
\maketitle
\flushbottom

\section{Introduction}
Squeezed states play a prominent role in the description of a wide range of phenomena including non-classical states of light in quantum optics \cite{walls2007quantum}, the emission of radiation from evaporating black holes \cite{Hawking:1974sw} and the amplification of quantum fluctuations into macroscopic perturbations during cosmic inflation \cite{mukhanov2005physical,parker2009quantum,Grishchuk:1990bj,Albrecht:1992kf,Polarski:1995jg}. In this paper we consider squeezed states for a system of $N$ bosonic degrees of freedom. The system can represent for instance the lattice discretization of a scalar field. Given a set of creation and annihilation operators $a_i$, $a^\dagger_i$ localized at each site of the lattice, we have a notion of Fock vacuum and a family of squeezed vacua defined as the set of states obtained by applying the squeeze operator $\exp(\frac{1}{2}\gamma^{ij}a_i^\dagger a_j^\dagger)$ to the Fock vacuum \cite{walls2007quantum}. In fact each squeezed vacuum can be understood as the Fock vacuum associated to some set $\tilde{a}_i$, $\tilde{a}^\dagger_i$ of creation and annihilation operators obtained from the first set via a Bogolyubov transformation \cite{berezin:1966}. The mathematical structure behind this ambiguity is the choice of a complex structure $J$: a squeezed vacuum $|J\rangle$ is labeled by an element of the space of positive compatible complex structures of the system. We derive a new formula for the entanglement entropy of a squeezed state: given a decomposition of the Hilbert space of the system in a tensor product $\mathcal{H}_V=\mathcal{H}_A\otimes \mathcal{H}_B$, the entanglement entropy $S_A(|J\rangle)$ of a squeezed vacuum restricted to the subsystem $\mathcal{H}_A$ is
\begin{equation*}
S_A(|J\rangle) =\text{tr}\Big(\frac{\mathds{1}-P_A\,\text{i} J P_A}{2}\log\Big|\frac{\mathds{1}-P_A\,\text{i} J P_A}{2}\Big|\Big)\,,
\label{eq:}
\end{equation*}
where $P_A$ is the projector to the phase space $A$ of the subsystem and the trace is over $N$ dimensional matrices. This formula has a clear interpretation: there is entanglement between $A$ and $B$ only if the projection $P_A\,J P_A$ of the complex structure is not a complex structure for the subsystem. In the derivation we use in a crucial way the map between the space of complex structures and the coset $Sp(2N,\mathbb{R})/\text{\emph{Isot}}(J_0)$ of symplectic transformation that \emph{squeeze} $J_0$. This map allows us to formulate squeezed vacua as group-theoretical coherent states for the symplectic group \cite{perelomov:2012} and to determine their entanglement Hamiltonian.\\

The ground state of a free scalar field discretized on a lattice provides a simple example of squeezed vacuum. Its entanglement entropy has been computed in the classical works \cite{sorkin1983entropy,bombelli1986quantum,srednicki1993entropy} and shown to result in an area law \cite{eisert2010colloquium}. More recently upper and lower bounds on the area scaling have been determined using results on block banded matrices \cite{plenio2005entropy,cramer2006entanglement,cramer2006correlations,schuch2006quantum}. As an application of our formula, in Sec.~\ref{sec:area} we determine the entanglement entropy of the ground state of a free scalar field Hamiltonian discretized on a generic lattice and find the expression $S_A(|J\rangle)=\frac{c^4}{4}\big(\!\log\frac{1}{c}\big)\;\text{Area}(\partial A)\;+O(c^4)$ in the limit of small speed of sound $c$.

The states we are interested in form a larger class than just ground states of quadratic Hamiltonians with local interactions. Squeezed vacua can be equivalently characterized in terms of their $2$-point correlation function: in the Schr\"odinger representation they are complex Gaussian functions of $N$ variables. Gaussian states play a central role in quantum information theory as they constitute a versatile resource for quantum communication protocols \cite{holevo2012quantum}. Their entanglement entropy was derived in \cite{holevo1999capacity} and later formulated in terms of symplectic invariants \cite{Adesso:2007tx}. The analysis of this paper largely relies on the methods developed in this setting (see \cite{adesso:2014co,braunstein2005quantum,ferraro2005gaussian,weedbrook2012gaussian} for recent reviews). The focus here however is not on the covariance matrix but on the complex structure $J$. Our motivation comes from the possible field-theoretical applications of our formula. 

The complex structure plays a central role in the definition of the notion of particles in quantum field theory in curved space \cite{wald1994quantum,ashtekar1975quantum,ashtekar1980curiosity,ashtekar1980geometrical} and in the description of unitarily inequivalent representations of the canonical commutation relations \cite{berezin:1966,shale1962linear}. In the case of black hole evaporation the paradigmatic setting involves an initial state prepared in the in-vacuum $|J_{\text{in}}\rangle$, its evolution in a black hole space-time and its analysis in terms of out-particles defined by the complex structure $J_\text{out}$, \cite{Hawking:1974sw,wald1994quantum,ashtekar1975quantum,ashtekar1980curiosity,ashtekar1980geometrical}. At late times the evolution of the state $|J_{\text{in}}\rangle$ is a squeezed vacuum with respect to the complex structure $J_\text{out}$. Its entanglement entropy and thermal properties have mostly been studied using conformal field theory techniques in a two-dimensional approximation \cite{Holzhey:1994we,Bianchi:2014qua,Bianchi:2014bma}. In the case of inflationary cosmology \cite{mukhanov2005physical,parker2009quantum}, an initial state $|J_{BD}\rangle$ evolves in a quasi de Sitter spacetime. The roll-over transition results in an instability in the evolution Hamiltonian and at a cosmic time $\eta$ before the end of inflation the state evolves in a highly squeezed vacuum $|J_\eta\rangle$,  \cite{mukhanov2005physical,parker2009quantum,Grishchuk:1990bj,Albrecht:1992kf,Polarski:1995jgAgullo:2015qqa}. In this paper we do not address directly the question of the definition and  computation of the entanglement entropy of such squeezed states in quantum field theory: we restrict attention to systems with finitely many degrees of freedom, effectively introducing a lattice cut-off of the continuum theory. In this setting we study the evolution of a squeezed state in the presence of an instability. The instability is modeled using a quadratic lattice Hamiltonian that is unbounded from below. In the long time limit we find the remarkably simple result reported in Sec.~\ref{sec:instability}: the rate of growth of the entanglement entropy is asymptotically bounded from above by the Kolmogorov-Sinai rate of the associated classical system \cite{zaslavsky2008hamiltonian} and is only mildly dependent on the initial squeezed state and the choice of subsystem $A$.\\

We begin by defining the structures involved in the derivation of our main formula for the entanglement entropy $S_A(|J\rangle)$ of a squeezed vacuum $|J\rangle$. The main formula is derived in Sec.~\ref{sec:entropy} and used in section Sec.~\ref{sec:field} to study the area law and the time-evolution of the entanglement entropy in the presence of instabilities.

\section{Symplectic group and $N$ bosonic degrees of freedom}
In this section we summarize some basic results about the symplectic group and its unitary representation on the Hilbert space of a system of $N$ bosonic degrees of freedom \cite{arnold-givental:1990,berndt:2001,degosson:2006,folland:1989}.

\subsection{Symplectic vector space $(V,\Omega)$ and the space of complex structures $\mathcal{J}(V,\Omega)$}
Consider a linear vector space $V=\mathbb{R}^{2N}$ of finite dimension $2N$. We denote $v^a$ the elements of $V$ and $w_a$ the elements of its linear dual $V\,\hat{}$. A symplectic structure on $V$ is a bilinear map $\Omega_{ab}:V\times V\to \mathbb{R}$ that is antisymmetric
\begin{equation}
\Omega_{ab}\, v^a u^b=-\Omega_{ab}\, u^a v^b
\label{eq:}
\end{equation}
and non-degenerate, i.e.
\begin{equation}
\Omega_{ab}\,v^a u^b=0\quad \forall u^a\in V\quad \Rightarrow \quad v^a=0\,.
\label{eq:}
\end{equation}
The inverse is denoted $\Omega^{ab}: V\,\hat{}\times V\,\hat{}\to \mathbb{R}$ and satisfies the relation $\Omega^{ac}\,\Omega_{bc}={\delta^a}_b$, where ${\delta^a}_b$ is the identity. The couple $(V,\Omega_{ab})$ is a symplectic vector space of dimension $2N$. 

A linear subspace $A\subset V$ is a symplectic subspace of $V$ if the intersection with its symplectic complement
\begin{equation}
A^{\scriptscriptstyle \Omega}=\{v^a\in V|\,\Omega_{ab}v^a u^b=0,\;\,\forall\, u^b\in A\}
\label{eq:symp-compl}
\end{equation}
is null, i.e. $A\cap A^{\scriptscriptstyle \Omega}=\{0\}$. Defining $B\equiv A^{\scriptscriptstyle \Omega}$ we have the decomposition $V=A\oplus B$. We denote $P_A:V\to V$ the projector from $V$ to the subspace $A$.

A linear symplectic transformation is a linear map ${M^a}_{b}:V\to V$ that preserves the symplectic structure
\begin{equation}
\Omega_{ab}\;{M^a}_{c}\,{M^b}_{d}\,=\,\Omega_{cd}\,,
\label{eq:}
\end{equation}
or equivalently in terms of the inverse symplectic structure ${M^a}_{c}\,{M^b}_{d}\,\Omega^{cd}=\,\Omega^{ab}$. Linear symplectic transformations ${M^a}_{b}$ form a group, the linear symplectic group $Sp(2N,\mathbb{R})$.\\

On the symplectic vector space $V$ we introduce a further structure, a linear complex structure, i.e. a linear map ${J^a}_{b}:V\to V$ that squares to minus the identity,
\begin{equation}
{J^a}_{b} \,{J^b}_{c}=-{\delta^a}_c\,.
\label{eq:}
\end{equation}
The complex structure ${J^a}_{b}$ is said to be compatible to the symplectic structure $\Omega_{ab}$ and positive if the following two conditions are satisfied:
\begin{align}
\Omega_{ab}\;{J^a}_{c}\,{J^b}_{d}\,&\,=\,\Omega_{cd}\,, \label{eq:cs1}\\[.5em]
\Omega_{ab}\,{J^b}_c&\,>\,0\,.\label{eq:cs2}
\end{align}
The first equation tells us that the complex structure belongs to the linear symplectic group, ${J^a}_b\in Sp(2N,\mathbb{R})$, the second that together with the symplectic structure it defines a metric, i.e. a positive definite inner product $g_{ab}:V\times V\to\mathbb{R}$ given by
\begin{equation}
g_{ab}\,\equiv\, \Omega_{ac}\,{J^c}_b\,.
\label{eq:metricOJ}
\end{equation}

We are interested in the set of all complex structures on $(V,\Omega)$ that are compatible and positive. We denote this set $\mathcal{J}(V,\Omega)$,
\begin{equation}
\mathcal{J}(V,\Omega)=\{{J^a}_b\,|\, \text{positive compatible complex structure on}\; (V,\Omega)\}\,.
\label{eq:}
\end{equation}
Note that the projection to $A$ of a complex structure, i.e. $P_A J P_A$, in general does not induce a complex structure on $A$.

\subsection{Polar decomposition and the $\text{\emph{Squeeze}}(J_0)$ coset}
In the following we introduce a matrix notation for symplectic transformations and their polar decomposition. It is evident from its definition that this notation involves the choice of a reference complex structure $J_0$. 

We denote a vector $v^a\in V$ simply by $v\equiv (v^a)$. Similarly a linear map ${M^a}_b: V\to V$ is denoted $M\equiv ({M^a}_b)$. The metric $g_{0\,ab}\,\equiv\, \Omega_{ac}\,J_0{{}^c}_b$ can be understood as a linear map from the vector space to its dual, $g_{0\,ab}:V\to V\,\hat{}$. We use this map to lower indices, i.e. $u_a\equiv g_{0\,ab}\,u^b$, and define the transpose $u^t\equiv (g_{0\,ab}\,u^b)$.\footnote{This notation is consistent with $\Omega^{ab}$ being the inverse of $\Omega_{ab}$.}  

Note that the definition of the transpose involves the choice of a reference complex structure $J_0\equiv (J_0{{}^a}_b)$. For definiteness we choose a reference complex structure $J_0$ associated to a decomposition of the symplectic vector space $V=\mathbb{R}^{2N}$ in symplectic subspaces $V_i=\mathbb{R}^{2}$ so that $V=\oplus_i V_i$ with $i=1,\ldots, N$. The restriction $\Omega_i$ of the symplectic structure $\Omega$ to each subspace $V_i$ can be put in the canonical form with an appropriate choice of coordinates. Given a positive compatible complex structure $j_0$ in each $V_i$ we define $J_0$ as the direct sum over $i$. In particular we choose
\begin{equation}
j_0=\Big(
\begin{array}{cc}
0 \,&\! -1\\[-2pt]
1\,&\,0
\end{array}
\Big)\,,\qquad
J_0=\bigoplus_{i=1}^N j_0\;\;=\;\; \Bigg(
\begin{array}{cc}
\,0\; &\! -\mathds{1}\\[4pt]
\,\mathds{1}\;&0
\end{array}
\Bigg)\,,
\label{eq:J0plus}
\end{equation}
where $\mathds{1}=({\delta^i}_{\!j})$ is the $N\times N$ identity. It is immediate to check that the two conditions (\ref{eq:cs1}) and (\ref{eq:cs2}) are satisfied so that the complex structure is compatible with the symplectic structure and positive. With this choice of complex structure, the associated metric (\ref{eq:metricOJ}) is simply the Euclidean metric $g_{0\,ab}\equiv\Omega_{ac}\,J_0{}^c{}_b=\delta_{ab}\,$. In particular in the matrix notation discussed above the transpose of a vector is simply $u^t\equiv (\delta_{ab}\,u^b)$.\\

In matrix language, the complex structure $J_0$ satisfies ${J_0}^2=-\mathds{1}$ and ${J_0}^t=-J_0$. The linear symplectic group is the group of $2N\times 2N$ matrices satisfying $M J_0\, M^t=J_0$, i.e.
\begin{equation}
Sp(2N,\mathbb{R})=\{M\in\text{Mat}(2N,\mathbb{R})\,|\,M J_0\, M^t=J_0\}\,.
\label{eq:Sp-def}
\end{equation}
From the definition it is immediate to prove that a symplectic matrix $M$ is invertible and it has inverse\footnote{It is less immediate to prove that $\det M=+1$. An algebraic proof makes use of the notion of Pfaffian.} 
\begin{equation}
M^{-1}=-J_0\, M^t J_0\,.
\label{eq:inverseM}
\end{equation}

The linear symplectic group  $Sp(2N,\mathbb{R})$  is non-compact. A maximally compact subgroup can be identified with the set of elements of $Sp(2N,\mathbb{R})$ that leave the chosen complex structure  $J_0\in \mathcal{J}(V,\Omega)$ invariant under conjugation, i.e. the isotropy group $\text{\emph{Isot}}(J_0)$ defined as 
\begin{equation}
\text{\emph{Isot}}(J_0)=\{M\in Sp(2N,\mathbb{R})\,|\,M J_0\, M^{-1}=J_0\}\,.
\label{eq:}
\end{equation} 
Using (\ref{eq:inverseM}) we see that the elements of $\text{\emph{Isot}}(J_0)$ satisfy $M M^t=\mathds{1}$, i.e. they are orthogonal transformations in $O(2N,\mathbb{R})$ and
\begin{equation}
\text{\emph{Isot}}(J_0)=Sp(2N,\mathbb{R})\cap O(2N,\mathbb{R})\,.
\label{eq:}
\end{equation}
As $O(2N,\mathbb{R})$ is a compact group, it follows that also the isotropy group is compact, in fact it is a maximally compact subgroup of $Sp(2N,\mathbb{R})$.\\

Symplectic transformations that `squeeze' the complex structure $J_0$, i.e. that do not leave it invariant under conjugation, form the coset space
\begin{equation}
\text{\emph{Squeeze}}(J_0)= Sp(2N,\mathbb{R})/\text{\emph{Isot}}(J_0)\,.
\label{eq:}
\end{equation}
The polar decomposition allows us to write uniquely every symplectic matrix as a product of a matrix $R\in \text{\emph{Isot}}(J_0)$ and a matrix $T\in \text{\emph{Squeeze}}(J_0)$,
\begin{equation}
M=T\,R\,,
\label{eq:M=TR}
\end{equation}
where $T=(M\, M^t)^{\frac{1}{2}}$ and  $R=(M\, M^t)^{-\frac{1}{2}}M$. It is clear that $R R^t=\mathds{1}$ and therefore $R\in  O(2N,\mathbb{R})$. The matrix $T=(M\, M^t)^{\frac{1}{2}}$ is symmetric and positive, $T\in \text{Sym}_+(2N,\mathbb{R})$ and provides a choice of representative in the squeeze coset
\begin{equation}
\text{\emph{Squeeze}}(J_0)\simeq Sp(2N,\mathbb{R})\cap \text{Sym}_+(2N,\mathbb{R})\,.
\label{eq:}
\end{equation}
The Lie algebra $\mathfrak{sp}(2N,\mathbb{R})$ of the symplectic group and its decomposition in generators of $\text{\emph{Isot}}(J_0)$ and of $\text{\emph{Squeeze}}(J_0)$ is discussed in App.~\ref{App:algebra}.

The space of positive compatible complex structures $\mathcal{J}(V,\Omega)$ can naturally be identified with the $\text{\emph{Squeeze}}(J_0)$ coset,
\begin{equation}
\mathcal{J}(V,\Omega)\simeq \text{\emph{Squeeze}}(J_0)\,.
\label{eq:}
\end{equation}
In fact every $J\in \mathcal{J}(V,\Omega)$ can be written as $J=MJ_0 M^{-1}$ for some $M\in Sp(2N,\mathbb{R})$ and $\text{\emph{Squeeze}}(J_0)$ is the set of symplectic transformations that act non-trivially on $J_0$ under conjugation.

\subsection{Phase space and Poisson brackets}\label{sec:Poisson}
The symplectic vector space $(V,\Omega)$ can be understood as the phase space of a classical dynamical system with $N$ degrees of freedom. Consider coordinates $\xi^a$ on $V=\mathbb{R}^{2N}$ and smooth functions $f, g\in C^\infty(\mathbb{R}^{2N})$. The Poisson bracket 
\begin{equation}
\{f,g\}=\Omega^{ab}\,\partial_a f\,\partial_b g
\label{eq:}
\end{equation}
endows the space of smooth functions with a Lie algebra structure. 

The space of constant or linear functions on $V$, i.e. functions of the form $f(\xi^a)=c+z_a\xi^a$, forms a sub-algebra called the Heisenberg algebra $\mathfrak{h}_{2N+1}$.  Canonical variables are linear functions $q^i=q^i_a\xi^a$, $p^i=p^i_a\xi^a$ such that they satisfy the canonical Poisson brackets $\{q^i,p^j\}=\delta^{ij}$, $\{q^i,q^j\}=0$, $\{p^i,p^j\}=0$ with $i,j=1,\ldots, N$. In more compact form, by viewing the coordinates $\xi^a$ as functions on phase space, we can write the canonical Poisson brackets as
\begin{equation}
\{\xi^a,\xi^b\}=\Omega^{ab}\,.
\label{eq:}
\end{equation}
Linear canonical transformations are transformations $\xi^a\to {M^a}_b\,\xi^b$ that preserve the canonical Poisson brackets, i.e. ${M^a}_c\,{M^b}_d\,\Omega^{cd}=\,\Omega^{ab}$. Therefore the matrix ${M^a}_b$ is an elements of the symplectic group $Sp(2N,\mathbb{R})$.

The space of quadratic functions on $V$, i.e. functions of the form
\begin{equation}
\textstyle H=\frac{1}{2}K_{ab}\,\xi^a\,\xi^b\,,
\label{eq:}
\end{equation}
also forms a subalgebra:
\begin{equation}
\{H_1,H_2\}=H_3
\label{eq:}
\end{equation}
where $H_3=\Omega^{cd}K_{1\,ac}K_{2\,bd}\;\xi^a\,\xi^b$. This algebra is isomorphic to the Lie algebra $\mathfrak{sp}(2N,\mathbb{R})$ of the symplectic group (App.~\ref{App:algebra}). Quadratic functions on phase space are the generators of linear canonical transformations.

\subsection{Quantization, the Hilbert space $\mathcal{H}_V$, creation and annihilation operators}\label{sec:annihilation}
Quantization of the symplectic vector space $(V,\Omega)$ results in a Hilbert space $\mathcal{H}_V$ of $N$ bosonic degrees of freedom. The elementary observables in the quantum theory satisfy the canonical commutation relations 
\begin{equation}
[\xi^a,\xi^b]=\text{i}\,\Omega^{ab}\,.
\label{eq:CCRxi}
\end{equation}
The Hilbert space $\mathcal{H}_V$ provides a regular representation of the Heisenberg algebra $\mathfrak{h}_{2N+1}$ and can be realized in the Schr\"odinger representation as the space $L^2(\mathbb{R}^N)$ of square-integrable functions depending on a maximally commuting set of linear observables, e.g. $\psi(q^i)$. For finite $N$, the Stone-von Neumann theorem states that irreducible representations of the canonical commutation relations on a Hilbert space are unitarily equivalent to the Schr\"odinger representation \cite{rosenberg:2004}. In the following we adopt the Fock representation defined in terms of creation and annihilation operators \cite{berezin:1966}.

The reference complex structure $J_0$ defined in Eq.~(\ref{eq:J0plus}) can be diagonalized 
in $V_{\mathbb{C}}=V\otimes \mathbb{C}=\mathbb{C}^{2N}$ with a unitary transformation\footnote{Note that $\text{i}J_0$ is a Hermitian matrix.} and put in the form
\begin{equation}
J_0=\text{i}\sum_{i=1}^{N} (u_i\,u_i^\dagger-u^*_i\,u_i^{*\,\dagger})
\label{eq:}
\end{equation}
where $u_i$ and $u_i^*$ are orthonormal eigenvectors in $\mathbb{C}^{2N}$, with 
\begin{equation}
u_i^a=\frac{1}{\sqrt{2}}(\delta_i^a-\text{i}\,\delta_{i+N}^a)\,.
\label{eq:}
\end{equation}
The vectors $u_i^a$ span the eigenspace ${V_\mathbb{C}}^{\!+}\subset V_\mathbb{C}$ corresponding to the eigenvalue $+\text{i}$ of $J_0$, and analogously $u_i^{a\,*}$ spans the eigenspace ${V_\mathbb{C}}^{\!-}$ with eigenvalue $-\text{i}$. Creation and annihilation operators associated to the complex structure $J_0$ are defined as
\begin{equation}
a_i=u_{ia}^*\,\xi^a\;,\qquad a^\dagger_i=u_{ia}\,\xi^a\,.
\label{eq:ai}
\end{equation}
Using the expression $\mathds{1}=\sum_i (u_i\,u_i^\dagger+u^*_i\,u_i^{*\,\dagger})$ for the identity we can write the operator $\xi^a$ as 
\begin{equation}
\xi^b=\sum_{i=1}^{N} (u_i^b\, a_i\,+\,u^{*b}_i\,a_i^\dagger)\,.
\label{eq:}
\end{equation}
Thanks to the compatibility of $J_0$ to the symplectic structure, the creation and annihilation operators satisfy the canonical commutation relations
\begin{equation}
[a_i,a_j^\dagger]=\delta_{ij}\;,\qquad [a_i,a_j]=0\;,\qquad [a_i^\dagger,a_j^\dagger]=0\,.
\label{eq:}
\end{equation}
The Hilbert space $\mathcal{H}_V$ is defined algebraically by introducing the reference vacuum $|J_0\rangle$ defined as the state annihilated by all the $a_i$
\begin{equation}
a_i|J_0\rangle=0\qquad i=1,\ldots, N
\label{eq:aJ}
\end{equation}
and the orthonormal set of states 
\begin{equation}
|n_1,\ldots,n_N;J_0\rangle=\Big(\prod_{i=1}^N\frac{(a_i^\dagger)^{n_i}}{\sqrt{n_i!}}\Big)|J_0\rangle\,,
\label{eq:onbasis}
\end{equation}
with $n_i\in \mathbb{N}_0$. The Hilbert space $\mathcal{H}_V$ is the completion of the span of the Fock basis $|n_i;J_0\rangle\equiv|n_1,\ldots,n_N;J_0\rangle$.  A generic state $|s\rangle\in \mathcal{H}$ can be expanded on the basis (\ref{eq:onbasis}) and is given by
\begin{equation}
|s\rangle=\sum_{n_i}c_{n_i}\,|n_i;J_0\rangle\,,
\label{eq:}
\end{equation}
with $c_{n_i}$ a complex square-summable sequence.\\

The relation between the complex structure $J_0$ and the state $|J_0\rangle$ can be made manifest by noticing that the matrix
\begin{equation}
\frac{\mathds{1}-\text{i}J_0}{2}=\sum_{i=1}^{N} u_i\,u_i^\dagger
\label{eq:}
\end{equation}
is the projector from $V_\mathbb{C}={V_\mathbb{C}}^{\!+}\oplus {V_\mathbb{C}}^{\!-}$ to ${V_\mathbb{C}}^{\!+}$. Therefore Eq.~(\ref{eq:aJ}) can be written as
\begin{equation}
\frac{\mathds{1}-\text{i}J_0}{2}\xi\;|J_0\rangle=0\,.
\label{eq:}
\end{equation}
Note that a different choice of complex structure $J$ in $\mathcal{J}(V,\Omega)$ corresponds to a different choice of state $|J\rangle$ that can be used as reference vacuum for the Fock basis. The state $|J\rangle$ is defined by 
\begin{equation}
\frac{\mathds{1}-\text{i}J}{2}\xi\;|J\rangle=0\,.
\label{eq:}
\end{equation}
analogously to the vacuum $|J_0\rangle$ associated to the complex structure $J_0$.

\subsection{Unitary representation of $Sp(2N,\mathbb{R})$ on the Hilbert space $\mathcal{H}_V$}\label{sec:EF}
The Hilbert space $\mathcal{H}_V$ carries a unitary representation of the symplectic group $Sp(2N,\mathbb{R})$,\footnote{The unitary operators $U(M)$ provide in fact a projective representation of the symplectic group. A double covering of the group is required to make the representation single valued. This double covering is called the metaplectic group. An example of the two-valued nature of the representation is the following. Consider the element $M_t=\exp(t J_0)$ of $Sp(2N,\mathbb{R})$. We have $M_0=M_{2\pi}=\mathds{1}$. On the other hand the unitary representation of $M_t$ is given by $U(M_t)=\exp(\text{i}\,t\,\delta_{ab}\xi^a\xi^b)$ with the operator $\delta_{ab}\xi^a\xi^b$ having a half-integer spectrum for $N$ odd. As a result, while $U(M_0)=\mathds{1}$ we can have $U(M_{2\pi})|s\rangle=\pm|s\rangle$ on eigenstates of $\delta_{ab}\xi^a\xi^b$.} \cite{folland:1989,degosson:2006}
\begin{equation}
U(M)\,\xi^a\, U(M)^{-1}={M^a}_b\,\xi^b\,.
\label{eq:UxiU}
\end{equation}
Using the canonical commutation relations it can be shown that a quadratic hermitian operator $\frac{1}{2}K_{ab}\xi^a \xi^b$ with $K_{ab}\in\text{Sym}(2N,\mathbb{R})$ generates the unitary transformation representing a symplectic transformation $M$ with generator $\Omega^{ac}K_{cb}$, i.e. 
\begin{equation}
U(M)=\exp \big(\!{\textstyle-i \,\frac{1}{2}K_{ab}\xi^a \xi^b }\big)\,\qquad \text{with}\quad {M^a}_b=\exp(\Omega^{ac}K_{cb})\,.
\label{eq:expKJ}
\end{equation}
Symplectic transformation of the creation and annihilation operators associated to the reference complex structure $J_0$ follow from Eq.~(\ref{eq:ai}) and are given by
\begin{equation}
U(M)\,a_i\, U(M)^{-1}=\Phi_{ij}\,a_j+\Psi_{ij}\,a^\dagger_j
\label{eq:}
\end{equation}
with the Bogolyubov coefficients given by $\Phi_{ij}=u^\dagger_i M u_j$ and $\Psi_{ij}=u^\dagger_i M u^*_j$, \footnote{\label{fn:Mbogolyubov} This relation can be inverted so that the symplectic transformation $M$ is written in terms of the Bogolyubov coefficients as $M=\sum_{ij}(\Phi_{ij}u_i u_j^\dagger+\Psi_{ij}u_i u_j^{*\dagger}+\Psi^*_{ij}u^*_i u_j^{\dagger}+\Phi^*_{ij}u^*_i u_j^{*\dagger})$.} \cite{berezin:1966}. The set of quadratic hermitian operator $\frac{1}{2}K_{ab}\xi^a \xi^b$ that generate these unitary transformations can also be expressed in terms of creation and annihilation operators: we define the operators $E_{ij}$, $F_{ij}$ and $F^\dagger_{ij}$
\begin{align}
& E_{ij}=\frac{1}{2}(a_i^\dagger a_j+a_j a_i^\dagger)\,,\\[4pt]
& F_{ij}=a_i\, a_j\,,\quad
 F^\dagger_{ij}=a_i^\dagger\, a^\dagger_j\,.
\label{eq:}
\end{align} 
Their commutation relations provide a representation of the symplectic algebra $\mathfrak{sp}(2N,\mathbb{R})$,
\begin{align}
& [E_{ij},E_{kl}]=\delta_{ik}E_{jl}-\delta_{jl}E_{ik}\label{eq:uN}\\[4pt]
& [F_{ij},F_{kl}]=0\;,\qquad [F^\dagger_{ij},F^\dagger_{kl}]=0\\[4pt]
& [F_{ij},E_{kl}]=F_{il}\delta_{jk}+F_{jl}\delta_{ik}\\[4pt]
& [F^\dagger_{ij},E_{kl}]=-F^\dagger_{ik}\delta_{jl}-F^\dagger_{jk}\delta_{il}\\[4pt]
& [F_{ij},F^\dagger_{kl}]=E_{ik}\delta_{jl}+E_{il}\delta_{jk}+E_{jk}\delta_{il}+E_{jl}\delta_{ik}
\end{align}
From Eq.~(\ref{eq:uN}) we see that the $E_{ij}$ form a subalgebra: they are the generators of the unitary group $U(N)$, i.e.
\begin{equation}
\exp(\text{i} \alpha^{ij}\; E_{ij})\,a_i \,\exp(-\text{i} \alpha^{ij}\; E_{ij})=\Xi_{ij}\, a_j
\label{eq:UN}
\end{equation}
with $\Xi_{ij}$ a unitary matrix. The transformations generated by $E_{ij}$ do not mix the creation and annihilation operators $a_i^\dagger$ and $a_i$, therefore they preserve $J_0$ and are the generators of $\text{\emph{Isot}}(J_0)$.

Transformations generated by $F_{ij}$ and $F_{ij}^\dagger$ 
\begin{equation}
U=\exp\Big(\,\frac{1}{2}\beta^{ij}\,F_{ij}^\dagger\;-\frac{1}{2}\beta^{*\,ij}\,F_{ij}\Big)
\label{eq:Ubeta}
\end{equation}
necessarily mix creation and annihilation operators, $U \,a\,U^{-1}=\Phi \,a +\Psi\,a^\dagger$, and provide a representation of the $\text{\emph{Squeeze}}(J_0)$ symplectic transformations. The corresponding Bogolyubov coefficients are
\begin{equation}
\Phi=\cosh \sqrt{\beta\beta^\dagger}\,,\qquad\Psi=\frac{\sinh \sqrt{\beta\beta^\dagger}}{\sqrt{\beta\beta^\dagger}}\;\beta\,.
\label{eq:}
\end{equation}

%Polar decomposition of the squeezing parameter $\beta=\rho\, e^{\text{i}\theta}$ where $\rho=(\beta\beta^*)^{\frac{1}{2}}$ and $e^{\text{i}\theta}=(\beta\beta^*)^{-\frac{1}{2}}\,\beta$. Note $\rho=\rho^\dagger$, $\theta=\theta^\dagger$.
%\begin{equation}
%\Phi=(\cosh \rho)\,,\qquad \Psi=(\sinh \rho)\;e^{\text{i}\theta}
%\label{eq:}
%\end{equation}

A convenient parametrization of $\text{\emph{Squeeze}}(J_0)$ is provided by the Siegel domain $\mathcal{S}$ of complex symmetric matrices $\gamma$ in the unit disk \cite{berndt:2001}, i.e.
\begin{equation}
\mathcal{S}=\{\gamma\in \text{Mat}(N,\mathbb{C})|\,\gamma=\gamma^t\;\;\text{and}\quad 1-\gamma\gamma^\dagger>0\,\}\,.
\label{eq:}
\end{equation}
The squeeze unitary transformation (\ref{eq:Ubeta}) can be put in the form
\begin{align}
U=&\exp\Big(\,\frac{1}{2}\beta^{ij}\,F_{ij}^\dagger\;-\frac{1}{2}\beta^{*\,ij}\,F_{ij}\Big)\\
=&\exp\Big(\,\frac{1}{2}\gamma^{ij}\,F_{ij}^\dagger\Big)\exp\Big(\frac{1}{2}\log(1-\gamma\gamma^\dagger)^{ij}\,E_{ij}\Big)\exp\Big(\,\frac{1}{2}\gamma^{*ij}\,F_{ij}\Big)
\label{eq:Ubetagamma}
\end{align}
where we have defined the matrix
\begin{equation}
\gamma=\frac{\tanh \sqrt{\beta\beta^\dagger}}{\sqrt{\beta\beta^\dagger}}\;\beta\,.
\label{eq:}
\end{equation}
Note that $\gamma$ is symmetric and satisfies $0\leq \gamma\gamma^\dagger\leq\mathds{1}$ because of the properties of the hyperbolic tangent, therefore it belongs to the Siegel disk $\mathcal{S}$. We can write the Bogoliubov coefficients in terms of $\gamma$,
\begin{equation}
\Phi=(1-\gamma\gamma^\dagger)^{-\frac{1}{2}}\,,\qquad \Psi=(1-\gamma\gamma^\dagger)^{-\frac{1}{2}}\;\gamma\,.
\label{eq:}
\end{equation}
Note that $\gamma=\Phi^{-1}\Psi$. Using the expression for the symplectic transformation in terms of the Bogolyubov coefficients,  
\begin{equation}
M_\gamma=\sum_{ij}(\Phi_{ij}u_i u_j^\dagger+\Psi_{ij}u_i u_j^{*\dagger}+\Psi^*_{ij}u^*_i u_j^{\dagger}+\Phi^*_{ij}u^*_i u_j^{*\dagger})
\label{eq:}
\end{equation}
we can also write the symplectic matrix $M$ in terms of $\gamma$. By definition the reference complex structure $J_0$ transforms non-trivially under conjugation by $M_{\gamma}$, it is squeezed. We label a complex structure with a matrix $\gamma$ and define
\begin{equation}
J_\gamma=M_\gamma \,J_0\,M_\gamma^{-1}\,.
\label{eq:}
\end{equation}
The space of complex structures $\mathcal{J}(V,\Omega)$ can therefore be described by a reference complex structure $J_0$ and a point $\gamma$ in the Siegel disk $\mathcal{S}$.

\section{Squeezed vacuum states: properties}
Squeezed vacuum states can be characterized in a variety of equivalent ways. Here we follow the approach of \cite{perelomov:2012} and show that squeezed vacua are group-theoretical coherent states for the symplectic group. In particular we show that they provide a resolution of the identity for the even sector of the Hilbert space of the system. Moreover we discuss the squeezing of their correlations.

\subsection{Vacuum state associated to a complex structure}
A squeezed vacuum associated to the complex structure $J\in \mathcal{J}(M,\Omega)$ is a state in the Hilbert space $\mathcal{H}_V$ that satisfies the defining condition
\begin{equation}
\frac{\mathds{1}-\text{i}J}{2}\xi\;|J\rangle=0\,.
\label{eq:}
\end{equation}
Analogously to what we have done for the reference complex structure $J_0$ we can introduce a set $\tilde{u}_i$ of orthonormal eigenvectors with eigenvalue $+\text{i}$, i.e. $J \tilde{u}_i=+\text{i}\,\tilde{u}_i$. The associated annihilation operator is $\tilde{a}_i=\tilde{u}_{i a}^*\,\xi^a$ and the squeezed vacuum satisfies $\tilde{a}_i |J\rangle=0$. Using the parametrization $J_\gamma$ in terms of a point $\gamma$ in the Siegel disk  we can also write
\begin{equation}
(a_i+\gamma_{ij} a_j^\dagger)\,|J_\gamma\rangle=0\,,
\label{eq:}
\end{equation}
where $a_i$ and $a^\dagger_i$ are annihilation and creation operators associated to the reference complex structure $J_0$. Using Eq.~(\ref{eq:UxiU}) we see that the state $|J_\gamma\rangle$ can be obtained by squeezing the vacuum $|J_0\rangle$ with a unitary transformation $U(M_\gamma)$,
\begin{equation}
|J_\gamma\rangle=U(M_\gamma)\,|J_0\rangle\,.
\label{eq:}
\end{equation}
Using Eq.~(\ref{eq:Ubetagamma}) together with $E_{ij}|J_0\rangle=\frac{1}{2}\delta_{ij}|J_0\rangle$ and $F_{ij}\,|J_0\rangle=0$, we find the expression
\begin{equation}
|J_\gamma\rangle\;=\;\det(\mathds{1}-\gamma\gamma^\dagger)^{\frac{1}{4}}\,\exp\Big({\textstyle \frac{1}{2}}\gamma^{ij}F^\dagger_{ij}\Big)\,|J_0\rangle\,.
\label{eq:}
\end{equation}
This expression allows us to compute the scalar product of two squeezed vacua
\begin{equation}
\langle J_{\gamma_1}|J_{\gamma_2}\rangle=\left(\frac{\det(\mathds{1}-\gamma_1\gamma_1^\dagger)\,\det(\mathds{1}-\gamma_2\gamma_2^\dagger)}{\det(\mathds{1}-\gamma_1^\dagger\gamma_2)^2}\right)^{\frac{1}{4}}\,.
\label{eq:}
\end{equation}
In particular $\langle J_0|J_\gamma\rangle=\det(\mathds{1}-\gamma\gamma^\dagger)^{\frac{1}{4}}$. Moreover we can expand a squeezed vacuum over the basis (\ref{eq:onbasis}),
\begin{equation}
|J_\gamma\rangle=\sum_{n_i}c_{n_i}(\gamma)\,|n_i;J_0\rangle\,,
\label{eq:}
\end{equation}
with coefficients
\begin{equation}
c_{n_i}(\gamma)=\frac{\det(\mathds{1}-\gamma\gamma^\dagger)^{\frac{1}{4}}}{\pi^N}\int_{\mathbb{C}^N}e^{-\frac{1}{2}\sum\gamma_{ij}z_i z_j}\,\prod_{i=1}^N\left(\frac{(z_i^*)^{n_i}}{\sqrt{n_i!}}\;e^{-z_i z_i^*}dz_i\right)\,.
\label{eq:}
\end{equation}
Note that the coefficients $c_{n_1\cdots n_N}(\gamma)$ with $\sum_{i=1}^Nn_i$ odd all vanish: squeezed vacuum states have support on the sector of the Hilbert space spanned by Fock space basis elements $|n_i;J_0\rangle$ with $\sum_{i=1}^Nn_i$ even.

\subsection{Squeezed vacua as symplectic coherent states}
Group-theoretical coherent states are defined in terms of a Lie group $G$ and its unitary irreducible representation $U(g)$ on a Hilbert space $\mathcal{H}$, \cite{perelomov:2012}. Given a reference state $|v_0\rangle$ in $\mathcal{H}$, a set of states is obtained by applying a group transformation, $|v_g\rangle=U(g)|v_0\rangle$. A set of coherent states is obtained by an appropriate choice of the reference state. 

Let $\mathfrak{g}$ be the algebra of the group $G$, $\mathfrak{g}_\mathbb{C}$ its complex hull and $X_k$ the set of generators of $\mathfrak{g}_\mathbb{C}$. A generator $X_k$ preserves or annihilates the state $|v_0\rangle$ if there exists a $\lambda_k\in \mathbb{C}$ such that
\begin{equation}
X_k |v_0\rangle=\lambda_k \, |v_0\rangle\,.
\label{eq:maximal}
\end{equation}
Within the Hilbert space $\mathcal{H}$ there is a maximal set $\{X_1,\ldots,X_n\}$ of generators such that at least one solution of Eq.~(\ref{eq:maximal}) exists. The reference state $|v_0\rangle$ for the construction of coherent states is required to be one of these solutions.\footnote{This condition guaranties that the set of coherent states have minimal dispersions of the Casimir operators of the group \cite{perelomov:2012}.} A case of particular interest occurs when a subset of the generators $X_k$ forms a subalgebra $\mathfrak{h}$ of $\mathfrak{g}_\mathbb{C}$. In this case the reference state satisfies
\begin{equation}
U(g)|v_0\rangle=e^{\text{i}\theta}|v_0\rangle\;\;\;\forall\,g\in \text{\emph{Isot}}(h_0)\,.
\label{eq:}
\end{equation}
i.e. it is invariant up to a phase under the action of the isotropy group $\text{\emph{Isot}}(h_0)=\{g\in G|\,g\,h_0\,g^{-1}=h_0\}$ for some group element $h_0$. In this case a coherent state is labeled by a reference state $|v_0\rangle$ and an element of the coset $G/\text{\emph{Isot}}(h_0)$. This construction can be immediately applied to the unitary representation of $Sp(2N,\mathbb{R})$ discussed in the previous section to obtain symplectic coherent states of a bosonic lattice.\\

The Hilbert space $\mathcal{H}_V$ of our bosonic system provides a reducible unitary representation of the double-cover of the symplectic group $Sp(2N,\mathbb{R})$. Defining the parity operator $\Pi=\exp[2\pi\,\text{i}\,\sum_i (E_{ii}-\frac{1}{2})]$ we can decompose $\mathcal{H}_V$ in
\begin{equation}
\mathcal{H}_V=\mathcal{H}_{\text{even}}\oplus \mathcal{H}_{\text{odd}}\,.
\label{eq:}
\end{equation}
where $\mathcal{H}_{\text{even}}$ is the space of eigenstates of $\Pi$ with eigenvalue $+1$ and  $\mathcal{H}_{\text{odd}}$ with eigenvalue $-1$.\footnote{The sector $\mathcal{H}_{\text{even}}$ is spanned by the basis elements (\ref{eq:onbasis}) with $n_1+\cdots+ n_N$ even, $\mathcal{H}_{\text{odd}}$ by $n_1+\cdots+ n_N$ odd.}  As the unitary representation $U(M)=\exp \big(\!{\textstyle-i \,\frac{1}{2}K_{ab}\xi^a \xi^b }\big)$ is constructed out of quadratic operators it leaves the parity of the state invariant and $\mathcal{H}_V$ is therefore reducible.  An irreducible representation is induced on the even and the odd sector of the Hilbert space. 

Within $\mathcal{H}_{\text{even}}$ there is a unique state $|J_0\rangle$ that satisfies the condition (\ref{eq:maximal}) for the maximal set of generators $E_{ij}$ and $F_{ij}$,
\begin{equation}
E_{ij}|J_0\rangle=\frac{1}{2}\delta_{ij}|J_0\rangle\,,\quad \text{and}\qquad F_{ij}\,|J_0\rangle=0\,.
\label{eq:EgFg}
\end{equation}
This reference state is invariant up to a phase under the action of the isotropy group of the element element $J_0$ of $Sp(2N,\mathbb{R})$. A symplectic coherent state in $\mathcal{H}_{\text{even}}$ is obtained by acting on $|J_0\rangle$ with a symplectic transformation $U(M)$, i.e.
\begin{equation}
|J_M\rangle\equiv U(M)|J_0\rangle\,.
\label{eq:}
\end{equation}
Because of the invariance of the state $|J_0\rangle$ under the action of $\text{\emph{Isot}}(J_0)$, we can label a symplectic coherent state by the state $|J_0\rangle$ and an element of the squeeze coset $\text{\emph{Squeeze}}(J_0)=Sp(2N,\mathbb{R})/\text{\emph{Isot}}(J_0)$. This is exactly the squeezed vacuum state labeled by the complex structure $J=MJ_0 M^{-1}$.

Within $\mathcal{H}_{\text{odd}}$ there is no state that satisfies $E_{ij}|v_0\rangle=\lambda_{ij}|v_0\rangle$ for all $i,j$. There is however a set of $N$ orthogonal states that satisfy $F_{ij}|v_0\rangle=0$. Defining the single-excitation state $|1_k;J_0\rangle\equiv a_k^\dagger |J_0\rangle$ we have
\begin{equation}
F_{ij}\,|1_k;g_0\rangle=0\,,
\label{eq:}
\end{equation}
with $k=1\ldots N$. By choosing one of them as reference state, we have the set of symplectic coherent states  in $\mathcal{H}_{\text{odd}}$ given by
\begin{equation}
|1_k; J_M\rangle\equiv U(M)\, a^\dagger_k\,|J_0\rangle=\tilde{a}^\dagger_k\,|J_M\rangle\,,
\label{eq:}
\end{equation}
where $\tilde{a}^\dagger_k\equiv U(M) a^\dagger_k U(M)^{-1}$.

A remarkable property of group-theoretical coherent states is that they provide a resolution of the identity as in integral over the group. For symplectic coherent states we have
\begin{equation}
\mathds{1}=\int_{Sp(2N,\mathbb{R})}  |J_M\rangle \langle J_M|\;\,d\mu(M)\;\;\;+\;\;\;\int_{Sp(2N,\mathbb{R})}  |1_k; J_M\rangle \langle 1_k; J_M|\;\,d\mu(M)\,,
\label{eq:resolution}
\end{equation}
where $d\mu(M)$ is the Haar measure on $Sp(2N,\mathbb{R})$. In Eq.~(\ref{eq:resolution}) the first term projects on $\mathcal{H}_{\text{even}}$, the second on $\mathcal{H}_{\text{odd}}$. Symplectic coherent states provide an overcomplete basis of $\mathcal{H}$ and every state can be written as a linear superposition of $|J_M\rangle$ and $|1_1; J_M\rangle$ for instance. Clearly the integration over $M$ in Eq.~(\ref{eq:resolution}) is redundant: elements of the compact subgroup $\text{\emph{Isot}}(J_0)\sim U(N)$ leave the integrand $|J_M\rangle \langle J_M|$ invariant. We can parametrize the integral over the $\text{\emph{Squeeze}}(J_0)$ coset by using the Bloch-Messiah decomposition: we write $M=R\, T_0\, R'$ with $R$, $R'$ in $\text{\emph{Isot}}(J_0)$ and $T_0$ diagonal and in $\text{\emph{Squeeze}}(J_0)$. In particular we have
\begin{equation}
U(T_0)=\exp\Big(\sum_i s_i\, (F^\dagger_{ii}-F_{ii})\Big)
\label{eq:Us}
\end{equation}
with squeezing parameters $s_i\geq 0$ and $T_0=e^{X_0}$ with $X_0=$ $\text{Diag}(s_1,\ldots, s_N,$  $-s_1,\ldots,-s_N)$. Following \cite{lupo:2012} we find the induced invariant measure
\begin{equation}
\textstyle d\mu(T_0)=\big(\prod_{i<j=1}^N|\cosh s_i\,-\cosh s_j|\big)\;\prod_i ds_i\,.
\label{eq:}
\end{equation}
The following expressions provide equivalent representations of the resolution of the identity in $\mathcal{H}_{\text{even}}$:
\begin{align}
\mathds{1}_{\text{even}}&=\int_{\mathcal{J}(\Omega,V)}  |J\rangle \langle J|\;\,d\mu(J)\;=\;\int_{Sp(2N,\mathbb{R})}  |J_M\rangle \langle J_M|\;\,d\mu(M)\\[5pt]
&=\int_{U(N)\times \mathbb{R}^N}  |J_{RT_0}\rangle \langle J_{RT_0}| \;d\mu(R) d\mu(T_0)
\label{eq:}
\end{align}
where $d\mu(J)$ can be understood as the invariant measure on the space of positive compatible complex structures (or over the Siegel disk $\mathcal{S}$) and $d\mu(R)$ is the invariant measure on the unitary group $U(N)$.

\subsection{Correlations and uncertainty relations}
The expectation value of the elementary operator $\xi^a$ on a squeezed vacuum $|J\rangle$ is always vanishing,
\begin{equation}
\langle J|\xi^a|J\rangle=0\,.
\label{eq:}
\end{equation}
This is an immediate consequence of the fact that $\xi^a|J\rangle$ belongs to $\mathcal{H}_{\text{odd}}$ and is therefore orthogonal to $|J\rangle$. On the other hand the two-point correlation function is non-trivial and completely determined by the complex structure $J$,
\begin{equation}
C^{ab}\equiv\langle J|\xi^a\xi^b|J\rangle=\text{i}\,\Big(\frac{\mathds{1}-\text{i}J}{2}\Big)^a{}_c\;\Omega^{cb}\,.
\label{eq:Cab}
\end{equation}
This relation can be derived for instance by noticing that $\langle J_0|\xi^a\xi^b|J_0\rangle=\frac{1}{2}(\delta^{ab}+\text{i}\,\Omega^{ab})$ where $\delta^{ab}= J_0^a{}_c\,\Omega^{cb}$ and $\langle J|\xi^a\xi^b|J\rangle=M^a{}_c\,M^b{}_d\,\langle J_0|\xi^a\xi^b|J_0\rangle$.

Note that the set of vectors $v_a$ in the vector space $V\,\hat{}=\mathbb{R}^{2N}$ satisfying the condition $C^{ab}v_a v_b=1$ form an ellipsoid. In fact only the symmetric part of the correlation function determines the ellipsoid. This part coincides with half of the inverse metric $g^{ab}={J^a}_c\,\Omega^{cb}$ introduced in Eq.~(\ref{eq:metricOJ}), so that the correlation ellipsoid is defined by
\begin{equation}
\frac{1}{2}{J^a}_c\,\Omega^{cb}\,v_a v_b\,=\,1
\label{eq:}
\end{equation}
Using the Bloch-Messiah parametrization of $J=M J_0 M^{-1}$ with $M=R\, T_0$ we find that the principal axis of the ellipsoid come in pairs given by the vectors $R^a{}_i$ and $R^a{}_{i+N}$ that span a symplectic plane. The axis have length $e^{\pm s_i}$ where $s_i$ are the squeeze parameters introduced in Eq.~(\ref{eq:Us}). The ellipsoid associated to the state $|J_0\rangle$ is a sphere of radius $1$, i.e. $s_i=0$: squeezing the state $|J_0\rangle$ with a unitary transformation $U(M)$ produces a squeezed vacuum $|J\rangle$ with ellipsoid obtained by squeezing the sphere with the same symplectic transformation $M$.

Consider a pair of observables $q=q_a\xi^a$ and $p=p_a \xi^a$ in $\mathcal{H}_V$ that are canonically conjugate,\footnote{These observables correspond to a symplectic basis in a two-dimensional symplectic subspace of $(V,\Omega)$, i.e. $\Omega^{ab}q_a p_b=1$.} i.e. $[q,p]=\text{i}$. The dispersion\footnote{The dispersion of an observable $\mathcal{O}$ on a state $|s\rangle$ is defined as $\Delta \mathcal{O}\equiv\sqrt{\langle s| \mathcal{O}^2 |s\rangle-\langle s| \mathcal{O} |s\rangle\langle s| \mathcal{O} |s\rangle}$.} of $q$ and $p$   on the squeezed vacuum $|J\rangle$ is
\begin{equation}
\textstyle \Delta q=\sqrt{\frac{1}{2}\,{J^a}_c\,\Omega^{cb}\,q_a q_b}\;,\qquad \Delta p=\sqrt{\frac{1}{2}\,{J^a}_c\,\Omega^{cb}\,p_a p_b}\,.
\label{eq:}
\end{equation}
Using the fact that ${J^a}_c\,\Omega^{cb}$ is a positive matrix of determinant $1$ when restricted to a symplectic subspace we can show that 
\begin{equation}
 \Delta q\; \Delta p\;=\frac{1}{2}\sqrt{1+({J^a}_c\,\Omega^{cb}\,q_a p_b)^2}\,.
\label{eq:}
\end{equation}
In particular, the Heisenberg uncertainty relations for the canonical couple $q,p$ are saturated only when  ${J^a}_c\,\Omega^{cb}\,q_a p_b=0$ and are squeezed otherwise.

\section{Entanglement spectrum  and entropy of squeezed vacua}\label{sec:entropy}
We are interested in describing properties of a subsystem. In terms of the symplectic vector space $(V,\Omega)$, we consider a subspace $A=\mathbb{R}^{2N_A}$ of dimension $2N_A$. Its symplectic complement $A^{\scriptscriptstyle \Omega}$ is defined in Eq.~(\ref{eq:symp-compl}). We require that $A$ is a symplectic vector space with respect to the induced symplectic structure, $(A,\Omega)$. This requirement corresponds to the condition $A\cap A^\Omega=\{0\}$. The vector space $B=A^\Omega$ is also symplectic. Therefore we have a symplectic decomposition of $V$,
\begin{equation}
V=A\oplus B\,,
\label{eq:V=A+B}
\end{equation}
with $B$ of dimension $2N_B$ and $N_A+N_B=N$. Without loss of generality, we assume that $A$ is the smallest of the two, $N_A\leq N_B$. We denote $P_A$ and $P_B: V\to V$ the projectors from $V$ to the subspaces $A$ and $B$ respectively.

At the level of the Hilbert space $\mathcal{H}_V$ the symplectic decomposition (\ref{eq:V=A+B}) corresponds to a tensor product structure
\begin{equation}
\mathcal{H}_V=\mathcal{H}_A\otimes \mathcal{H}_B\,.
\label{eq:}
\end{equation}

In this section we derive a Schmidt decomposition of symplectic matrices analogous to the one discussed in \cite{Adesso:2007tx} and write the reduced density matrix 
\begin{equation}
\rho_A\equiv\text{Tr}_B\big(|J\rangle\langle J|\big)
\label{eq:}
\end{equation}
of a squeezed vacuum in terms of the associated entanglement Hamiltonian $H_A$. The entanglement entropy of the state $|J\rangle$ restricted to the subsystem $A$ is the von Neumann entropy of the reduced density matrix, 
\begin{equation}
S_A(|J\rangle)\equiv-\text{Tr}_A\big(\rho_A \log \rho_A\big),
\label{eq:}
\end{equation}
where the trace is over the Hilbert space $\mathcal{H}_A$. The entanglement entropy can be easily computed from the spectrum of the entanglement Hamiltonian \cite{Li:2008kda}. Following \cite{holevo2012quantum} we show that, at given two-point correlation function, squeezed vacua are the states with the largest entanglement entropy, a special case of the more general result of \cite{wolf2006extremality}. The expression of the entanglement entropy in terms of the projection of the complex structure is derived in Sec.~\ref{sec:SJ}.

\subsection{Normal modes and decomposition of the complex structure}
Given a complex structure $J$ on $V$ we are interested in its projection to the symplectic subspace $A$, i.e. $P_A\, J P_A$. In general the projected complex structure does not induce the complex structure on the subspace. Here we use Williamson theorem \cite{arnold-givental:1990,berndt:2001,degosson:2006,Adesso:2007tx} to put the projection of a complex structure in canonical form.\\

Williamson theorem states that a positive-definite symmetric matrix $K\in\text{Sym}_+(2N,\mathbb{R})$ can be put in a canonical diagonal form via a symplectic transformation, i.e. $K= M (\Lambda\oplus \Lambda)M^t$ where $\Lambda=\text{Diag}(\lambda_1,\ldots,\lambda_N)$ with $\lambda_i>0$ and $M\in Sp(2N,\mathbb{R})$. The invariants $\lambda_i$ are the normal modes of $K$. We provide a constructive proof of this theorem.

Given $K\in\text{Sym}_+(2N,\mathbb{R})$ consider the hermitian matrix $H=\sqrt{K}\;\text{i}J_0\,\sqrt{K}$. As $H^\dagger=H$ and $H^*=-H$ we have that it has real eigenvalues with opposite sign associated to orthogonal eigenvectors that are complex conjugate, i.e.
\begin{equation}
H v_i=\lambda_i v_i\quad \text{and}\quad H v^*_i=-\lambda_i v^*_i\
\label{eq:}
\end{equation}
with $\lambda_i> 0$ and $i=1\ldots N$. The matrix $H$ is diagonalized by a $2N\times 2N$ unitary transformation $U=(v_1\cdots v_n \;v_1^*\cdots v_n^*)$
so that 
\begin{equation}
H=U\;(\Lambda\;\oplus\;-\Lambda)\,U^{-1}
\label{eq:}
\end{equation}
with $\Lambda=\text{Diag}(\lambda_1,\ldots,\lambda_N)$ and $(\Lambda\;\oplus\;-\Lambda)$ the block diagonal matrix with $\Lambda$ and $-\Lambda$ on the diagonal. Let us define the unitary matrix
\begin{equation}
W=\bigoplus_{i=1}^N \frac{1}{\sqrt{2}}\Big(
\begin{array}{cc}
1 \,&\, \text{i}\\[-2pt]
1\,&\!-\text{i}
\end{array}
\Big)\;\;= \frac{1}{\sqrt{2}}\left(
\begin{array}{cc}
\,\mathds{1}\, & \;\,\text{i}\,\mathds{1}\\[4pt]
\,\mathds{1}\, &-\text{i}\,\mathds{1}
\end{array}
\right)\,.
\label{eq:}
\end{equation}
Note that, because of the structure of the eigenvectors of $H$, the matrix $R=W^\dagger U^\dagger$ is real and orthogonal, $R R^t=\mathds{1}$. We define now the matrix $M$
\begin{equation}
M=\sqrt{K}\,U\, W\, (\sqrt{\Lambda^{-1}}\oplus \sqrt{\Lambda^{-1}}\,)\,.
\label{eq:williamsonM}
\end{equation}
It is immediate to check that $MJ_0M^t=J_0$, therefore $M\in Sp(2N,\mathbb{R})$. This concludes the constructive proof of Williamson theorem: a positive-definite symmetric matrix $K$ can be written in terms of the normal modes $\lambda_i$ as
\begin{equation}
K= M (\Lambda\oplus \Lambda)M^t
\label{eq:}
\end{equation}
where the invariants $\lambda_i$ are the absolute value of the eigenvalues of the hermitian matrix $H=\sqrt{K}\;\text{i}J_0\,\sqrt{K}$ and the symplectic matrix $M$ is given by Eq.~(\ref{eq:williamsonM}). We note also that the matrix $\sqrt{K} H \sqrt{K^{-1}}$ is diagonalizable and it has the same eigenvalues of the matrix  $H$. Therefore the Williamson invariants $\lambda_i$ can be directly computed as the absolute value of the eigenvalues of the matrix $\text{i}K J_0$.\\

\bigskip

%\subsection{Decomposition of the complex structure}

A complex structure $J\in \mathcal{J}(V,\Omega)$ can be obtained from the reference complex structure $J_0$ by conjugation with a symplectic transformation, $J=MJ_0 M^{-1}$. Using the polar decomposition (\ref{eq:M=TR}) we can write $M=T R$ and $J=T^2 J_0$, where $T^2\in \text{Sym}_+(2N,\mathbb{R})$. We adopt the notation $[P_A M P_A]:A\to A$ to denote the  full-rank matrix obtained from $P_A M P_A:V\to V$.\footnote{A basis of $V=A\oplus B$ consisting of the union of a basis of $A$ and a basis of $B$ can be chosen. In this case the matrix elements of $[P_A M P_A]$ coincide with the matrix elements of $M$ on a restricted range. We adopt this more abstract notation because it simplifies some algebraic manipulations when we use an orthonormal basis that is not adapted to the decomposition $A\oplus B$.} By construction, the projection $P_A J_0 P_A$ of the reference complex structure $J_0$ to the subspace $A$ is also a complex structure that we will still call $J_0=[P_A J_0 P_A]$. The projection of $J$ on the other hand does not belong in general to $\mathcal{J}(A,\Omega)$ and can be written as 
\begin{equation}
[P_A \,JP_A]=[P_A \,T^2P_A]\;J_0\,,
\label{eq:}
\end{equation}
with $[P_A \,T^2P_A]\in \text{Sym}_+(2N_A,\mathbb{R})$. We apply Williamson theorem to this matrix and we obtain
\begin{equation}
[P_A \,T^2P_A]=M_A \,(\Lambda\oplus \Lambda)\, M_A^t
\label{eq:}
\end{equation}
where $M_A\in \text{Sp}(2N_A,\mathbb{R})$ and $\Lambda=\text{Diag}(\cosh 2r_1,\ldots,\cosh 2r_{N_A})$. From Williamson theorem we know that the normal modes $\cosh 2r_i$ can be determined as the absolute value of the eigenvalues of $[P_A \,T^2P_A] \,\text{i}J_0$. This is exactly $\text{i}$ times the projection of the complex structure, so that we have $[P_A \,\text{i}JP_A]=\,M_A\,(\Lambda\oplus -\Lambda)\,M_A^{-1}$, i.e.
\begin{equation}
\text{Eigenvalues}([P_A \,\text{i}JP_A])=\{\pm\cosh 2r_i\}\,.
\label{eq:EvPA}
\end{equation} 
It follows that
\begin{equation}
\text{Eigenvalues}\Big([P_A \frac{\mathds{1}-\text{i}J}{2}P_A]\Big)=\{(\cosh r_i)^2\,,\;-(\sinh r_i)^2\}\,.
\label{eq:EvP1-JP}
\end{equation} 
The projection of a complex structure is a complex structure only if the ``squeeze'' parameters $r_i$ are all vanishing.\\

The same analysis can be performed for the projection $P_B \,\text{i}JP_B$ to the subspace $B$. The projected complex structure has the same non-vanishing eigenvalues $\{\pm\cosh 2r_i\}$, plus $2(N_B-N_A)$ eigenvalues $\pm1$. In particular, as for the projection to $A$, we can write 
\begin{equation}
[P_B \,\text{i}JP_B]=\,M_B\,(\Lambda\oplus -\Lambda\;\oplus\; 1\oplus\,-1)\,M_B^{-1}
\label{eq:}
\end{equation}
where $M_B\in \text{Sp}(2N_B,\mathbb{R})$. When extended to $V$ we find that the symplectic transformation $M_A\oplus M_B\in \text{Sp}(2N,\mathbb{R})$ puts the complex structure in the canonical form
\begin{equation}
J=\;(M_A\oplus M_B)\;T_0 \,J_0\, T_0^{-1}\,(M^{-1}_A\oplus M^{-1}_B)
\label{eq:}
\end{equation}
with $T_0$ a symplectic transformation in $\text{\emph{Squeeze}}(J_0)$ given by
\begin{equation}
T_0=\exp\Big(\sum_{i=1}^{N_A}r_i\, K_i J_0\Big)
\label{eq:}
\end{equation}
with 
\begin{equation}
K_i=\left(
\begin{array}{c|c:ccc}
0 & \kappa_i&\;&0&\; \\ 
\hline
\kappa_i& 0& \; & 0&\\
\hdashline\\[-12pt]
0&0 & & 0& 
\end{array}
\right)\;,
\qquad
\kappa_i=\left(
\begin{array}{cccc}  
0 & \quad&\quad&\quad\\ 
\quad & 0&\quad&\quad\\
\quad & \quad&1&\quad\\
\quad & \quad&\quad&0\\
\end{array}
\right)\,,
\label{eq:}
\end{equation} 
or more explicitly
\begin{equation}
T_0=\left(
\begin{array}{c|c:ccc}
c\oplus c & s\oplus -s&\;&0&\; \\ 
\hline
s\oplus -s & c\oplus c& \; & 0&\\
\hdashline\\[-12pt]
0&0 & & \mathds{1} & 
\end{array}
\right)
\label{eq:}
\end{equation}
with $c=\text{Diag}(\cosh r_1, \ldots ,\cosh r_{N_A})$ and $s=\text{Diag}(\sinh r_1, \ldots ,\sinh r_{N_A})$.

\subsection{Schmidt decomposition and the entanglement Hamiltonian}
Using the decomposition
\begin{equation}
M=(M_A\oplus M_B) \,T_0\,R
\label{eq:}
\end{equation}
we can write the squeezed vacuum state associated to the complex structure $J=M J_0 M^{-1}$ in the form
\begin{align}
|J\rangle&=U(M)|J_0\rangle= U(M_A\oplus M_B)U(T_0)|J_0\rangle\\
&= U_A\, U_B\, \exp\big(\sum_{i=1}^{N_A} r_i (a_i^\dagger b_i^\dagger-a_i b_i)\big)|J_0\rangle\\
&= U_A\, U_B\, \prod_{i=1}^{N_A} \Bigg(\sum_{n=0}^\infty \frac{(\tanh r_i)^n}{\cosh r_i}\; \frac{(a_i^\dagger)^n}{\sqrt{n!}} \frac{(b_i^\dagger)^n}{\sqrt{n!}}\Bigg)|J_0\rangle
\label{eq:}
\end{align}
where $b_i\equiv a_{N_A+i}$. Defining 
\begin{equation}
|J_{A}\rangle=U(M_A)\bigotimes_{i=1}^{N_A}|j_0\rangle_i\quad,\quad |J_{B}\rangle=U(M_B)\bigotimes_{i=N_A+1}^{N_A+N_B}|j_0\rangle_i
\label{eq:}
\end{equation}
and $\tilde{a}_i=U_A\, a_i\, U_A^{-1}$, $\tilde{b}_i=U_B\, b_i\, U_B^{-1}$ we have the orthonormal basis in $\mathcal{H}_A$ and  $\mathcal{H}_B$ 
\begin{equation}
|n_1,\ldots,n_{N_A};J_{A}\rangle=\Big(\prod_{i=1}^{N_A}\frac{(\tilde{a}_i^\dagger)^{n_i}}{\sqrt{n_i!}}\Big)|J_{A}\rangle\;,\quad |n_1,\ldots,n_{N_B};J_{B}\rangle_2=\Big(\prod_{i=1}^{N_B}\frac{(\tilde{b}_i^\dagger)^{n_i}}{\sqrt{n_i!}}\Big)|J_{B}\rangle\,.
\label{eq:}
\end{equation}
This is a Schmidt basis for the squeezed vacuum $|J\rangle$, i.e.
\begin{equation}
|J\rangle=\sum_{n_1\cdots n_{N_A}=0}^\infty  \left(\prod_{i=1}^{N_A}\frac{(\tanh r_i)^{n_i}}{\cosh r_i}\right)\;|n_1,\ldots,n_{N_A};J_A\rangle\otimes |n_1,\ldots,n_{N_A};J_B\rangle\,.
\label{eq:}
\end{equation}

\bigskip

%\subsection{Reduced density matrix and the entanglement Hamiltonian}
The reduced density matrix $\rho_A$ can now be easily computed,
\begin{equation}
\rho_A=\text{Tr}_B\Big(|J\rangle\langle J|\Big)=\sum_{n_1\cdots n_{N_A}=0}^\infty  \left(\prod_{i=1}^{N_A}\frac{(\tanh r_i)^{n_i}}{\cosh r_i}\right)^2\;|n_1,\ldots,n_{N_A};J_A\rangle \langle n_1,\ldots,n_{N_A};J_A|\,.
\label{eq:}
\end{equation}
This expression can also be written as
\begin{equation}
\rho_A=\frac{e^{-2\pi H_A}}{Z_A}
\label{eq:rhoA}
\end{equation}
where $H_A$ is the entanglement Hamiltonian
\begin{equation}
H_A=-\frac{1}{\pi}\sum_{i=1}^{N_A}\log(\tanh r_i) \,\tilde{a}_i^\dagger\tilde{a}_i
\label{eq:}
\end{equation}
and $Z_A=\text{Tr}(e^{-H_A})=\prod_{i=1}^{N_A} (\cosh r_i)^2$. The squeezing parameters $r_i$ are obtained from the eigenvalues $\pm\cosh 2r_i$ of $\text{i}J$ restricted to region $A$, Eq.~(\ref{eq:EvPA}). The entanglement spectrum is the set of eigenvectors and eigenvalues of the entanglement Hamiltonian as computed above \cite{Li:2008kda}.

\subsection{Entanglement entropy and extremality of squeezed vacua}
The entanglement entropy of the squeezed vacuum $|J\rangle$ restricted to the sub Hilbert space $\mathcal{H}_A$ is given by the von Neumann entropy of the reduced density matrix,
\begin{equation}
S_A(|J\rangle)\equiv -\text{Tr}\big(\rho_A\log \rho_A\big)\,.
\label{eq:}
\end{equation}

Let us consider two squeezed vacua $|J_1\rangle$ and $|J_2\rangle$ on $\mathcal{H}_V$. For a given subsystem $A$ we want to compare the entanglement entropy of the two states. It is useful to define the correlation matrices $C^{ab}_1$ and $C^{ab}_2$ of the two states using Eq.~(\ref{eq:Cab}), together with the entanglement Hamiltonians appearing in the expression of the reduced density matrices $\rho_1$, $\rho_2$ of the two states,
\begin{equation}
H_1=\frac{1}{2}K_{1\,ab}\,\xi^a \xi^b \qquad H_2=\frac{1}{2}K_{2\,ab}\,\xi^a \xi^b\,.
\label{eq:}
\end{equation}
The relative entanglement entropy $R_A$ is the relative entropy of the reduced density matrices 
\begin{equation}
R_A(|J_2\rangle;|J_1\rangle)\equiv\text{Tr}\Big(\rho_2 \log\rho_2\,-\rho_2\log \rho_1\Big)\;\geq 0\,.
\label{eq:}
\end{equation}
It is always positive if the two density matrices do not coincide and zero if they do \cite{holevo2012quantum}. The difference between the entanglement entropies can be bounded from below using the following inequality
\begin{align}
S_A(|J_1\rangle)-S_A(|J_2\rangle)=&-\text{Tr}\Big(\rho_1\log \rho_1\Big)+\text{Tr}\Big(\rho_2\log \rho_1\Big)+R_A(|J_2\rangle;|J_1\rangle)\\
\geq& -\text{Tr}\Big(\rho_1\log \rho_1\Big)+\text{Tr}\Big(\rho_2\log \rho_1\Big)\\
&=\text{Tr}(H_1\,\rho_1)-\text{Tr}(H_1\,\rho_2)\\
&=\frac{1}{2}K_{1\,ab}\Big(\text{Tr}(\xi^a\xi^b\,\rho_1)-\text{Tr}(\xi^a\xi^b\,\rho_2)\Big)\\
&=\frac{1}{2}K_{1\,ab}\big(C_1^{ab}-C_2^{ab}\big)
\label{eq:}
\end{align}
Therefore a larger correlation function, $C_1^{ab}-C_2^{ab}\,>0$, implies a larger entropy for every subsystem $A$. In terms of the complex structures we have 
\begin{equation}
(J_1-J_2)^a{}_c\,\Omega^{cb}\,>0\quad\Rightarrow\quad S_A(|J_1\rangle)>S_A(|J_2\rangle)\,.
\label{eq:}
\end{equation}
The same method can be used to prove that squeezed vacuum states are the states with the largest entanglement entropy at fixed correlation function \cite{wolf2006extremality}. Consider a generic state $|s\rangle\in \mathcal{H}_V$ and define the connected correlation function of this state in the standard way
\begin{equation}
C_s^{ab}=\langle s|\xi^a \xi^b|s\rangle-\langle s|\xi^a |s\rangle\langle s|\xi^b|s\rangle\,.
\label{eq:}
\end{equation}
Out of $C_s^{ab}$ we can define a complex structure 
\begin{equation}
J^a{}_b=2\,C_s^{ac}\,\Omega_{cb}-\text{i}\,\delta^a{}_b\,.
\label{eq:}
\end{equation}
The associated squeezed vacuum $|J\rangle$ that has exactly the same correlation function $C_J^{ab}$ as the state $|s\rangle$. Then we have
\begin{align}
S_A(|J\rangle)-S_A(|s\rangle)=&-\text{Tr}\big(\rho_J\log \rho_J\big)+\text{Tr}\big(\rho_s\log \rho_J\big)+R_A(|s\rangle;|J\rangle)\\
&=\frac{1}{2}K_{ab}\Big(\text{Tr}(\xi^a\xi^b\,\rho_J)-\text{Tr}(\xi^a\xi^b\,\rho_s)\Big)+R_A(|s\rangle;|J\rangle)\\
&=\frac{1}{2}K_{ab}\big(C_J^{ab}-C_s^{ab}\big)+R_A(|s\rangle;|J\rangle)\\
&=R_A(|s\rangle;|J\rangle)\geq 0
\label{eq:}
\end{align}
Therefore at fixed correlation function the squeezed vacuum is the state with the largest entanglement entropy.

\subsection{Entanglement entropy of squeezed vacua in terms of the complex structure}\label{sec:SJ}
The entanglement entropy of the squeezed vacuum $|J\rangle$ restricted to the sub Hilbert space $H_A$ is given by $S_A(|J\rangle)\equiv -\text{Tr}\big(\rho_A\log \rho_A\big)$. 
%the von Neumann entropy of the reduced density matrix,
%\begin{equation}
%S_A(|J\rangle)\equiv -\text{Tr}\big(\rho_A\log \rho_A\big)\,.
%\label{eq:}
%\end{equation}
Using Eq.~(\ref{eq:rhoA}) for the reduced density matrix expressed in terms of the entanglement Hamiltonian $H_A$ it is immediate to show that
\begin{equation}
-\text{Tr}_A\big(\rho_A\log \rho_A\big)=\sum_{i=1}^{N_A} \big((\cosh r_i)^2\,\log (\cosh r_i)^2\;-\,(\sinh r_i)^2\,\log (\sinh r_i)^2\big)\,.
\label{eq:Sr}
\end{equation}
This expression can be cast in terms of the eigenvalues of $P_A \frac{\mathds{1}-\text{i}J}{2}P_A$, Eq.~(\ref{eq:EvP1-JP}). Equivalently, in matrix form, we find our final expression for the entanglement entropy in terms of the complex structure:
\begin{equation}
S_A(|J\rangle)=\text{tr}\Big(P_A \frac{\mathds{1}-\text{i}J}{2}P_A\,\log \Big| P_A \frac{\mathds{1}-\text{i}J}{2}P_A \Big|\Big)\,,
\label{eq:main}
\end{equation}
where the trace is over $2N$ dimensional matrices and not on the Hilbert space $\mathcal{H}_A$. This is the main formula studied in the rest of the paper.\\

Note that this formula is equivalent to the one found in \cite{holevo1999capacity} for the entropy of a Gaussian bosonic channel expressed in terms of its symplectic invariants, and the one of \cite{Sorkin:2012sn} for the entanglement entropy of a quasi-free state of a scalar field on a causal set, expressed in terms of the covariant correlation function. The expression in terms of the complex structure makes manifest the fact the entanglement entropy vanishes when $P_A \frac{\mathds{1}-\text{i}J}{2}P_A$ is a projector (in this case in fact its eigenvalues are either $0$ or $1$ and $0\log 0 =0$, $1\log 1 =0$). This happens if and only if the projection of the complex structure $P_A \,J P_A$ induces a complex structure for the subsystem $A$. To have a nonvanishing entanglement entropy we need that the complex structure $J$ sends a vector in $A$ into a vector with components in $B$. In fact, using $P_A+P_B=\mathds{1}$ and $(P_A \,\text{i}J P_A)^2=P_A+P_A \,J P_B\,J P_A$, we see that the eigenvalues of $P_A \,\text{i}J P_A$  are different from $\pm 1,0$ if and only if $P_B\,J P_A$ is non-vanishing.

\subsection{Entanglement entropy for small and large squeezing}
Suppose that the squeezing parameters $r_i$ in Eq.~(\ref{eq:EvPA}) are all small. This happens if $P_A \,JP_A$ is close to being a complex structure, in particular $\text{tr}\big((P_A \,\text{i}JP_A)^2-\mathds{1}\big)\ll N_A$. Expanding Eq.~(\ref{eq:Sr}) $r_i\ll 1$ we find
\begin{equation}
S_A(|J\rangle)= \sum_{i=1}^{N_A} (1-\log r_i^2)\, r_i^2\;+\;O(r_i^4)\,.
\label{eq:Srsmall}
\end{equation}
On the other hand for large squeezing $r_i\gg 1$, i.e. $\text{tr}\big((P_A \,\text{i}JP_A)^2-\mathds{1}\big)\gg N_A$, we find the asymptotic expansion
\begin{equation}
S_A(|J\rangle)\approx \Big(\sum_{i=1}^{N_A} 2r_i\Big)+(1-2\log 2)N_A\,.
\label{eq:Srlarge}
\end{equation}
The leading order can be written in matrix form in terms of the projected complex structure as
\begin{equation}
S_A(|J\rangle)\approx\frac{1}{2}\log\big|\!\det([P_A \,\text{i}JP_A])\big|\,,
\label{eq:Sdet}
\end{equation}
where $[P_A \,\text{i}JP_A]$ is the full rank matrix obtained from the projected symplectic structure.

\section{Scalar field on a lattice and squeezed vacua}\label{sec:field}
In this section we apply the formalism of squeezed vacua and complex structures to study the entanglement entropy of a scalar field on a lattice. The lattice provides a notion of locality for the interaction between the $N$ bosonic degrees of freedom. 

The entanglement entropy of the ground state of a local Hamiltonian \cite{sorkin1983entropy,bombelli1986quantum,srednicki1993entropy} is well-studied \cite{plenio2005entropy,cramer2006entanglement,cramer2006correlations,schuch2006quantum} and know to lead to an area law \cite{eisert2010colloquium}. In Sec.~\ref{sec:area} we consider a generic lattice and use formula Eq.~(\ref{eq:main}) to determine the dependence of the area law on the speed of sound, in the limit of  small speed of sound.

The time dependence of the entanglement entropy for a state that is not an eigenstate of the Hamiltonian has been studied using a variety of methods \cite{amico2004dynamics,Calabrese:2005in,AbajoArrastia:2010yt,Hartman:2013qma,van2013entanglement,Avery:2014dba}. For a chaotic system the rate of the entanglement entropy growth is expected to be related to the Lyapunov exponents of the system \cite{Zurek:1994wd,pattanayak1999lyapunov,petitjean2006lyapunov,Asplund:2015osa,Maldacena:2015waa}. In Sec.~\ref{sec:instability} we study the evolution of a squeezed vacuum in the presence of an instability described by a quadratic Hamiltonian with potential that is unbounded from below. We find a precise relation between the long-time behavior of the entanglement entropy, the size of the subsystem, the eigenvectors of the interaction potential and the eigenvalues associated to the unstable directions. In particular we find that the rate of growth of the entanglement entropy is asymptotically bounded from above by the Kolmogorov-Sinai rate of the associated classical system.

\subsection{Bosonic lattice: degrees of freedom and regions}\label{sec:lattice}
Consider a system of $N$ bosonic degrees of freedom $(\phi_i,\pi_i)$ with canonical commutation relations
\begin{equation}
[\phi_i,\pi_j]=\text{i}\,\delta_{ij}\;,\quad [\phi_i,\phi_j]=0\;,\quad [\pi_i,\pi_j]=0\,, \quad\qquad i,j=1,\ldots, N\,.
\label{eq:CCR}
\end{equation}
The canonical variables $(\phi_i,\pi_i)$ are defined by $\phi_i=\delta_{ia}\xi^a$ and $\pi_i=\delta_{i+N\,a}\,\xi^a$ in terms of the observable $\xi^a$ on $\mathcal{H}_V$. We can define a set of annihilation and creation operators as
\begin{equation}
a_i=\frac{1}{\sqrt{2}}(\phi_i+\text{i}\,\pi_i)\,,\qquad a^\dagger_i=\frac{1}{\sqrt{2}}(\phi_i-\text{i}\,\pi_i)\,.
\label{eq:annihilation}
\end{equation}
These are the operators associated to the reference complex structure $J_0=\oplus_{i=1}^N j_0$ introduced in Eq.~(\ref{eq:J0plus}). We denote $|J_0\rangle$ the vacuum annihilated by $a_i$ and $|n_i;J_0\rangle$ the Fock basis built over this vacuum as in Eq.~(\ref{eq:onbasis}).

In order to define a notion of locality and adjacency relations between the degrees of freedom we equip the system with a lattice structure. We introduce a graph $\Gamma$ with $N$ nodes: each node plays the role of site $i$ where the degree of freedom $(\phi_i,\pi_i)$ lives.\\

Consider the set $\mathcal{N}=(1,2,\ldots,N)$. We call \emph{nodes} the elements $i\in \mathcal{N}$ and we equip functions $v_i$ on this set with the structure of a vector space. The adjacency relation of two nodes $i,j$ is determined by the adjacency matrix $t_{ij}$ \cite{godsil:2013}, a linear map from  $\mathcal{N}$ to $\mathcal{N}$ defined by
\begin{equation}
t_{ij}=\left\{
\begin{array}{ll}
1 \qquad &   \text{if} \;i,j\; \text{distinct and adjacent}, \\
0  &     \text{otherwise}.
\end{array}
\right.
\label{eq:tij}
\end{equation}
We call \emph{link} the unordered couple $\ell_{(ij)}=(i,j)$ with $i,j$ adjacent nodes. We denote $\mathcal{L}=\{(i,j)|i,j\in \mathcal{N}, t_{ij}=1\}$ the set of links. The couple $\Gamma=(\mathcal{N},\mathcal{L})$ defines a finite graph that is undirected and simple. The number of links of the graph is determined by the adjacency matrix via the formula
\begin{equation}
L=\frac{1}{2}\sum_{i,j\in \mathcal{N}}t_{ij}\,.
\label{eq:}
\end{equation}
The valence $d_i$ of the node $i$ is defined as the number of links that contain the node $i$ and is given by
\begin{equation}
d_i=\sum_{j\in \mathcal{N}}t_{ij}\,.
\label{eq:}
\end{equation}
The adjacency matrix provides us also with a notion of graph Laplacian defined as
\begin{equation}
\Delta_{ij}=t_{ij}-d_i\, \delta_{ij}\,.
\label{eq:}
\end{equation}
The matrix $-\Delta_{ij}$ is positive semi-definite and its spectrum captures various properties of the graph. In particular in the following we assume that the graph $\Gamma$ is connected\footnote{A graph $\Gamma$ is connected if for every two nodes $i_1$, $i_M$ there is a sequence $(i_1,i_2,\ldots,i_M)$ such that $t_{i_1 i_2}t_{i_2 i_3}\cdots t_{i_{M-1} i_M}=1$.}. In this case there is at most one eigenvalue $0$ in the spectrum of the graph Laplacian with the corresponding eigenvector being the constant vector. If we impose boundary conditions on the eigenvectors, such as vanishing components on a set of nodes of the graph, then all eigenvalues are strictly positive.

We introduce now the notion of subgraph $\Gamma_A$. Consider a subset $\mathcal{N}_A\subset \mathcal{N}$ with $N_A$ nodes. We denote $\mathcal{N}_B$ its complement so that $\mathcal{N}=\mathcal{N}_A\cup\mathcal{N}_B\,$. We denote 
\begin{equation}
\big(\Pi_A\big)_{ij}=\sum_{k\in\mathcal{N}_A}\delta_{ki}\delta_{kj}
\label{eq:PiA}
\end{equation}
the projector from $\mathcal{N}$ to $\mathcal{N}_A$. We also define the subset $\mathcal{L}_A=\{(i,j)|i,j\in \mathcal{N}_A,\, t_{ij}=1\}\subset \mathcal{L}$ defined as the set of links connecting two nodes in $\mathcal{N}_A$, and the set of \emph{boundary links} 
\begin{equation}
\mathcal{L}_{\partial A}=\{(i,j)|i\in \mathcal{N}_A, \,j\in \mathcal{N}_B,\, t_{ij}=1\}\,.
\label{eq:}
\end{equation}
We have $\mathcal{L}=\mathcal{L}_A\cup\mathcal{L}_B\cup\mathcal{L}_{\partial A}$. The subgraph $\Gamma_A\subset \Gamma$ is defined as $\Gamma_A=(\mathcal{N}_A,\mathcal{L}_A)$. In general the subgraph $\Gamma_A$ can have more than one connected component.\\

A bosonic lattice $(\Gamma, \mathcal{H}_V)$ is defined by a graph $\Gamma$ with a bosonic degree of freedom $(\phi_i,\pi_i)$ at each node $i\in\mathcal{N}$ and the associated Hilbert space
\begin{equation}
\mathcal{H}_V=\bigotimes_{i\in \mathcal{N}}L^2(\mathbb{R})\,.
\label{eq:}
\end{equation}
A region of the bosonic lattice is a coucple $(\Gamma_A, \mathcal{H}_A)$ where $\Gamma_A$ is a connected subgraph of $\Gamma$ with the bosonic degree of freedom $(\phi_i,\pi_i)$ with $i\in\mathcal{N}_A$. The Hilbert space of the region $A$ is
\begin{equation}
\mathcal{H}_A=\bigotimes_{i\in \mathcal{N}_A}L^2(\mathbb{R})\,.
\label{eq:}
\end{equation}
Notice that we have the tensor product structure $\mathcal{H}_V=\mathcal{H}_A\otimes \mathcal{H}_B$. This decomposition corresponds to a decomposition of symplectic vector spaces $V=A\oplus B$, where $A=\mathbb{R}^{2N_A}$ is the vector space spanned by the canonical basis $\{e_i, \,J_0 e_i\}$ with $e_i=(\delta_i^a)$ and $i=1,\ldots,N_A$. The projector from $V$ to $A$ is
\begin{equation}
P_A=\sum_{i=1}^{N_A}\big(e_ie_i^t+(J_0e_i)(J_0e_i)^t\Big)\,.
\label{eq:}
\end{equation}
In terms of the projector $\Pi_A$ to the subset of sites $\mathcal{N}_A$, Eq.~(\ref{eq:PiA}), we have $P_A=\Pi_A\oplus \Pi_A$.

\subsection{Ground state of Hamiltonians and the area law}\label{sec:area}
Consider the lattice Hamiltonian
\begin{equation}
H=\sum_{i,j=1}^N\frac{1}{2}\Big(\delta_{ij} \pi_i\pi_j+V_{ij}\phi_i\phi_j\Big)\,
\label{eq:}
\end{equation}
with $V_{ij}\phi_i\phi_j$ a potential term bounded from below, i.e. $V=(V_{ij})\in \text{Sym}_+(N,\mathbb{R})$. The Hamiltonian can be put in normal form
\begin{equation}
H=\sum_{k=1}^N\omega_k\big(b^\dagger_k b_k+\frac{1}{2}\big)
\label{eq:}
\end{equation}
where $\omega_k$ are the eigenvalues of the matrix $\sqrt{V}$ and $b_k=u_{ka}\xi^a=\frac{1}{\sqrt{2}}\big(V^{1/4}_{\!ki}\phi_i+\text{i}\,V^{-1/4}_{\!ki}\pi_i\big)$ is the annihilation operator associated to the complex structure
\begin{equation}
J= \Bigg(
\begin{array}{cc}
\,0\; &\! -\sqrt{V^{-1}}\\[4pt]
\,\sqrt{V}\;&0
\end{array}
\Bigg)\,.
\label{eq:}
\end{equation}
This expression can be obtained by writing the Hamiltonian in the form $H=\frac{1}{2}K_{ab}\xi^a\xi^b$ and applying Williamson theorem to $K_{ab}$. The ground state of the Hamiltonian $H$ is the squeezed vacuum $|J\rangle$ that satisfies $b_k |J\rangle=0$.

The entanglement entropy $S_A(|J\rangle)$ can be obtained by computing the eigenvalues $\lambda=\pm \cosh 2r_i$ of the projected complex structure $P_A \,\text{i}JP_A$. The eigenvalues of $P_A \,\text{i}JP_A$ satisfy
\begin{align}
0\;&=\;\det\big(P_A \,\text{i}JP_A-\lambda\,\mathds{1}\big)\;=\;\det\Bigg(
\begin{array}{cc}
-\lambda\,\mathds{1} & -\text{i}\,\Pi_A\sqrt{V^{-1}}\Pi_A\\[4pt]
\text{i}\,\Pi_A\sqrt{V}\Pi_A& -\lambda\,\mathds{1}
\end{array}
\Bigg)\nonumber\\[8pt]
&=\;\det\Big(\lambda^2\,\mathds{1}-\Pi_A\sqrt{V}\,\Pi_A\sqrt{V^{-1}}\,\Pi_A\Big)\nonumber\\[8pt]
&=\;\det\Big(\Pi_A\sqrt{V}\,\Pi_B\sqrt{V^{-1}}\,\Pi_A\;-(1-\lambda^2)\,\mathds{1}\Big)\,,
\label{eq:}
\end{align}
where we have used $\Pi_A+\Pi_B=\mathds{1}$ in the last line. Therefore we find 
\begin{equation}
(\sinh 2r_i)^2=-\text{Eigenvalues}\Big(\Pi_A\sqrt{V}\,\Pi_B\sqrt{V^{-1}}\,\Pi_A\Big)\,,
\label{eq:s2r}
\end{equation}
The squeezing parameters $r_i$ are non-vanishing and the entanglement entropy non-trivial only if the potential $V_{ij}$ entangles $A$ with $B$, i.e. only if $\Pi_BV\,\Pi_A\neq 0$.\\

In the following we consider a special case: the lattice discretization of the scalar field Hamiltonian on a trivial background, i.e. the potential term
\begin{equation}
V_{ij}=\delta_{ij}-c^2\Delta_{ij}\,.
\label{eq:}
\end{equation}
This potential corresponds to setting the lattice spacing and the mass of the scalar field to unity, while keeping the speed of sound $c$ free. Note that the lattice is not assumed to be regular. It is instructive to study the behavior of the entanglement entropy for small speed of sound. For $c\ll1$ we have
\begin{equation}
-\Pi_A\sqrt{V}\,\Pi_B\sqrt{V^{-1}}\,\Pi_A=\frac{c^4}{4}\Pi_A\Delta\,\Pi_B\Delta\,\Pi_A\,+O(c^6)=\frac{c^4}{4}\Upsilon\,+O(c^6)
\label{eq:}
\end{equation}
where $\Upsilon= \Pi_At\,\Pi_Bt\,\Pi_A$ is the adjacency matrix $t_{ij}$ restricted to the boundary links $\mathcal{L}_{\partial A}$. As a result, using Eq.~(\ref{eq:s2r}),
\begin{equation}
\{r_i^2\}=\text{Eigenvalues}\Big(\frac{c^4}{16}\Upsilon\Big)\,+O(c^6)\,.
\label{eq:}
\end{equation}
Substituting in Eq.~(\ref{eq:Srsmall}) we find
\begin{align}
S_A(|J\rangle)&= \text{tr}\Big(\big(1-\log\frac{c^4}{16}\Upsilon\big)\frac{c^4}{16}\Upsilon\Big)\,+O(c^6)\\[8pt]
&=\frac{c^4}{4}\Big(\!\log\frac{1}{c}\Big)\;\text{tr}\,\Upsilon\;+O(c^4)\;=\;\frac{c^4}{4}\Big(\!\log\frac{1}{c}\Big)\;\text{Area}(\partial A)\;+O(c^4)
\label{eq:sound}
\end{align}
where $\text{tr}\,\Upsilon=\#\mathcal{L}_{\partial A}$ is the number of links of the graph that cross the boundary the region $A$ i.e., in a lattice discretization, the area of the boundary of the region $A$ in units of the lattice cut-off.

\subsection{Linear generation of entanglement at instabilities}\label{sec:instability}
We consider again a lattice Hamiltonian of the form
\begin{equation}
H=\sum_{i,j=1}^N\frac{1}{2}\Big(\delta_{ij} \pi_i\pi_j+V_{ij}\phi_i\phi_j\Big)\,
\label{eq:}
\end{equation}
but this time we do not assume that the Hamiltonian is bounded from below: the potential $V_{ij}\phi_i\phi_j$ term is assumed to have $N_I$ unstable directions. We write
\begin{equation}
V=\sum_{k=1}^{N_I}-\lambda^2_k\;v_k v_k^t\;+\;\sum_{k=N_I+1}^{N}\!\omega^2_k\;v_k v_k^t
\label{eq:Vinstab}
\end{equation}
where $v_k$ are eigenvectors of the potential, $-\lambda^2_k$ negative eigenvalues corresponding to the unstable directions and $\omega^2_k$ positive eigenvalues corresponding to the stable directions. We order the negative eigenvalues so that $\lambda_1\geq \lambda_2\geq \cdots\geq \lambda_{N_I}>0$. We assume also that the eigenvalues are non-degenerate. We are interested in the time evolution of the entanglement entropy of an initial state $|J_0\rangle$. By writing the Hamiltonian in the form $H=\frac{1}{2}K_{ab}\xi^a\xi^b$ we find that the time evolution operator
\begin{equation}
e^{-\text{i}H \,t}=U(e^{-\Omega Kt})
\label{eq:}
\end{equation}
corresponds to the symplectic transformation $M=\exp(-\Omega K\,t)$ where $\Omega K=(\Omega^{ac}K_{cb})$.\footnote{We have $K=\Bigg(
\begin{array}{cc}
\mathds{1} \;& 0\\[4pt]
0\;&V
\end{array}
\Bigg)$ 
and 
$\Omega K=\Bigg(
\begin{array}{cc}
0 \;& V\\[4pt]
-\mathds{1}\;&0
\end{array}
\Bigg)$.
} Therefore the evolution of the initial squeezed vacuum $|J_0\rangle$ is still a squeezed vacuum
\begin{equation}
|J_t\rangle= e^{-\text{i}H \,t}|J_0\rangle
\label{eq:}
\end{equation}
with 
\begin{equation}
J_t=e^{-\Omega K t}J_0\,e^{\Omega K t}\,.
\label{eq:}
\end{equation}
This expression in principle allows us to compute the entanglement entropy $S_A(|J_t\rangle)$ as a function of time. Here we study the long-time limit under the assumption that there is at least one unstable direction, $\lambda_1>0$.\\

The spectrum of $\Omega K$ can be expressed in terms of the eigenvalues and eigenvectors of the potential $V$. The eigenvalues $\mu$ of $\Omega K$ satisfy the equation\footnote{The first determinant is over $2N$-dimensional matrices while the second on $N$-dimensional matrices}
\begin{equation}
0=\det(\Omega K-\mu \mathds{1})=\det(V+\mu^2 \mathds{1}),
\label{eq:}
\end{equation}
therefore $\Omega K$ has $2N_I$ real eigenvalues $\mu=\pm \lambda_k$ and $2(N-N_I)$ imaginary eigenvalues $\mu=\pm\text{i} \omega_k$. The associated eigenvectors are $\frac{1}{\sqrt{2}}(\mp\lambda_k v_k,\, v_k)$ and $\frac{1}{\sqrt{2}}(\mp\text{i}\omega_k v_k,\, v_k)$. Using these results we can diagonalize the symplectic transformation $M=\exp(-\Omega K\,t)$, i.e.
\begin{equation}
M=N D N^{-1}
\label{eq:}
\end{equation}
with $D=\text{Diag}(e^{+\lambda_1 t}, e^{-\lambda_1 t},\ldots, e^{+\lambda_{N_I} t}, e^{-\lambda_{N_I} t}\;,\;e^{+\text{i}\omega_{N_I+1} t},e^{-\text{i}\omega_{N_I+1} t},\ldots,e^{+\text{i}\omega_{N} t},e^{-\text{i}\omega_{N} t})$ and\footnote{The inverse of $N$ is
\begin{equation}
N^{-1}=\frac{1}{\sqrt{2}}\left(
\begin{array}{cc}
-\lambda_1^{-1} v^t_1 \;&\;\; v_1^t \\[4pt]
+\lambda_1^{-1} v^t_1 \;&\;\; v_1^t \\[4pt]
\vdots & \vdots \\[4pt]
+\text{i}\omega_N^{-1} v^t_N \;&\;\; v_N^t \\[4pt]
-\text{i}\omega_N^{-1} v^t_N \;&\;\; v_N^t 
\end{array}
\right)\,.
\label{eq:}
\end{equation}}
\begin{equation}
N=\frac{1}{\sqrt{2}}\Bigg(
\begin{array}{ccccc}
-\lambda_1 v_1 \;& +\lambda_1 v_1&\cdots &-\text{i}\, \omega_N v_N \;& +\text{i}\,\omega_N v_N\\[4pt]
v_1\;&v_1&\cdots& v_N \;&  v_N
\end{array}
\Bigg)\,.
\label{eq:}
\end{equation}
Note that
\begin{equation}
M M^t=\sum_{k=1}^NC_k \,P_k 
\label{eq:MMt}
\end{equation}
where $P_k=v_k v_k^t$ is the projector on the eigenspace $\lambda_k$ of the potential, $C_k$ is the $2\times 2$ matrix (with $\det C_k=1$)
\begin{equation}
C_k=\frac{1}{2}\Bigg(
\begin{array}{cc}
1-\lambda_k^2+(1+\lambda_k^2)\cosh (2\lambda_k t) \;\,&\;  -(\lambda_k+\lambda_k^{-1})\sinh (2\lambda_k t) \\[8pt]
-(\lambda_k+\lambda_k^{-1})\sinh (2\lambda_k t) \;\; &\; 1-\lambda_k^{-2}+(1+\lambda_k^{-2})\cosh (2\lambda_k t) 
\end{array}
\Bigg)\,,
\label{eq:Ck2x2}
\end{equation}
and we have adopted the notation $\lambda_k\equiv \text{i}\,\omega_k$ for $k=N_I+1,\ldots,N$.\\

The entanglement entropy $S_A(|J_t\rangle)$ depends on the projected complex structure $P_A\, J_t P_A$, Eq.~(\ref{eq:main}). Under the assumption of large squeezing $r_i\gg 1$ (to be checked a posteriori), we can use the asymptotic expression (\ref{eq:Sdet}),
\begin{equation}
S_A(|J_t\rangle)\approx\frac{1}{2}\log\big|\!\det([P_A \,\text{i}J_t P_A])\big|\,.
\label{eq:SAJt}
\end{equation}
In the following we also assume, without loss of generality, that $A$ is the smallest of the two subsystems in the bipartition, i.e. $N_A\leq N_B$. Note that, using $J_t=MJ_0M^{-1}=MM^tJ_0$ and $[P_A J_0 P_A]=J_0$, we have
\begin{equation}
\big|\!\det([P_A \,\text{i}J_tP_A])\big|=\big|\!\det([P_A M M^t P_A]\,\,\text{i}J_0)\big|=\big|\!\det([P_A M M^t P_A])\big|\,.
\label{eq:}
\end{equation}
Furthermore, using Eq.~(\ref{eq:MMt}) and the spatial projector $\Pi_A$ (Eq.~(\ref{eq:PiA})), we obtain
\begin{equation}
\det([P_A \,\text{i}J_tP_A]) = \det\sum_{k=1}^N C_k\, \Pi_A P_k\Pi_A = \det\Big( \sum_{k=1}^N c_k\Big) \, ,
\label{eq:CkCh}
\end{equation}
where we defined $c_k=C_k\, \Pi_A P_k\Pi_A$. We are interested in the asymptotic behavior of the entanglement entropy for large $t$, which requires the determination of the leading contribution to (\ref{eq:CkCh}) in this limit. We write the entries of the matrices in the summand in (\ref{eq:CkCh}) explicitly as
\begin{equation}
[c_k]^{(i\alpha)(j\beta)} = [C_k]^{\alpha \beta} \,v_k^i v_k^j \, , \qquad  i,j=1,\dots,N_A \, , \quad \alpha,\beta=1,2.
\label{eq:}
\end{equation}
Introducing an index $r=2(i-1) + \alpha$, and noting that that the bi-index $(i,\alpha)$ is completely determined by $r$, we can switch to the single-index representation $[c_k]^{rs} = [c_k]^{(i\alpha)(j\beta)}$.

We recall now a standard result about the determinant of a sum of matrices as in Eq.~(\ref{eq:CkCh}). Using the definition of the determinant in terms of the Levi Civita tensors $\epsilon_{r_1\cdots r_{2N_A}}$ and the multilinearity of the determinant, we find
\begin{equation}
\det\left(\sum_{k=1}^N c_k \right)\,=\,\sum_{k_1=1}^N\cdots \sum_{k_{2N_A} =1}^N b_{(k_1\cdots k_{2N_A})}\,,
\label{eq:det-sum}
\end{equation}
where $b_{(k_1\cdots k_{2N_A})}$ is given by
\begin{equation}
b_{(k_1\cdots k_{2N_A})}\equiv\frac{1}{(2N_A)!}\sum_{r_1\cdots r_{2N_A}=1}^{2N_A}\sum_{r'_1\cdots r'_{2N_A}=1}^{2N_A}\epsilon_{r_1\cdots r_{2N_A}}\epsilon_{r'_1\cdots r'_{2N_A}}\;[c_{k_1}]^{r_1r'_1}\cdots [c_{k_{2N_A}}]^{r_{2N_A} r'_{2N_A}}\,.
\label{eq:Leibniz}
\end{equation}
In particular, for instance, $b_{(1\cdots 1)}=\det c_1$. Let us focus now on a single term $b_{(k_1\cdots k_{2N_a})}$ in Eq.~(\ref{eq:det-sum}). Note that, for $k=1,\ldots,N_I$, each entry $[C_k]^{\alpha \beta}$ of the matrix $C_k$ diverges as $e^{2\lambda_k t}$ for large $t$, while the determinant $\det C_k=1$ remains bounded. Since we are interested in the entropy for large $t$, it is sufficient to our purposes to keep track of the factors of $e^{2\lambda_k t}$ in the calculation of the determinant. Now, the explicit form of the matrix $[C_k]^{\alpha \beta}$ given in (\ref{eq:Ck2x2}) leads to several cancellations and simplifications in Eq.~(\ref{eq:Leibniz}) for $b_{(k_1\cdots k_{2N_a})}$. Consider for instance $b_{(k\,k\,\cdots)}$ where there appear only two identical $k_i$'s. Naively one could expect an asymptotic behavior $\propto e^{4\lambda_k t}$, but it is in fact not the case. The reason is a cancellation that follows from $\det C_k=1$. In the case of an odd number of identical $k$'s appear, e.g. $b_{(k\,k\,k\,\cdots)}$, the contribution is at most $e^{2\lambda_k t}$ for large $t$. As a result, in the large $t$ asymptotics of $\det([P_A \,\text{i}J_tP_A])$, each of the $N_I$ exponential factors $e^{2\lambda_k t}$ can appear at most once.

In general, not all unstable modes $k=1,\dots,N_I$ contribute to the entropy. We define
\begin{equation}
\delta^A_k=\left\{\begin{array}{ll}
0\quad& \text{if}\;\;\Pi_A P_k \Pi_A=0\,,\\[8pt]
1\quad& \text{else}\,,
\end{array}\right.
\qquad \text{and} \qquad
\delta^B_k=\left\{\begin{array}{ll}
0\quad& \text{if}\;\;\Pi_B P_k \Pi_B=0\,,\\[8pt]
1\quad& \text{else}\,.
\end{array}\right.
\label{eq:}
\end{equation}
In particular $\delta^A_k=0$ corresponds to an eigenvector $v_k$ that has its only non-vanishing components in $B$. In this case $\Pi_A P_k\Pi_A=0$, the corresponding $k$ can be omitted from the sum in (\ref{eq:CkCh}), and no factor of $e^{2\lambda_k t}$ contributes to $S_A(|J_t\rangle)$ in the asymptotic regime. On the other hand, if $v_k$ has support in $A$ only, i.e. $\delta^B_k=0$, we have a contribution that is not exponentially-diverging because of a cancellation of two terms that follows again from $\det C_k=1$. Summarizing, a factor $e^{2\lambda_k t}$  contributes to the asymptotics only if the eigenvector $v_k$ associated to an unstable direction has non-vanishing components both in $A$ and in $B$. Moreover each eigenvector with support in $A$ only, i.e. $\delta^B_k=0$, reduces by $2$ the total number of factors $e^{2\lambda_k t}$ that appear in the asymptotics. As a result, for large $t$, the asymptotic behavior is:
\begin{equation}
\left| \det([P_A \,\text{i}J_tP_A])\right| \sim  \exp\Big( 2\sum_{k=1}^{N_*}\delta^A_k\,\delta^B_k\, \lambda_k\;\,t\Big)\,,
\label{eq:}
\end{equation}
where $N_*$ is the smallest integer that solves
\begin{equation}
\sum_{k=1}^{N_*}\delta^A_k\,\delta^B_k\,=\, \min\Bigg(\sum_{k=1}^{N_I}\delta^A_k\,\delta^B_k\;\,,\;\;2(\sum_{k=1}^{N}\delta^B_k)-2N_B\Bigg)\,.
\label{eq:}
\end{equation}
The number of eigenvalues $\lambda_k$ contributing non-trivially to the asymptotics is $\sum_{k=1}^{N_*}\delta^A_k\,\delta^B_k$ and is always smaller or equal than the size $2N_A$ of the subsystem. The eigenvalue $\lambda_{N_*}$ is the smallest element of the sequence $\lambda_1> \lambda_2> \cdots> \lambda_{N_I}>0$ that contributes to the asymptotics. Using Eq.~(\ref{eq:SAJt}) we find that the long-time dependence of the entanglement entropy  in the presence of instabilities is 
\begin{equation}
S_A(|J_t\rangle)\sim  \Big(\sum_{k=1}^{N_*}\delta^A_k\,\delta^B_k\, \lambda_k\Big)\;t \, .
\label{eq:SAt}
\end{equation}

\begin{figure}[t]
${}$\\[2.5em]
\flushleft \hspace{2em} $S_A(|J_t\rangle)$\hspace{18.5em}$\frac{\partial}{\partial t} S_A(|J_t\rangle)$\\[-6em]
\centering
\includegraphics[height = 12em]{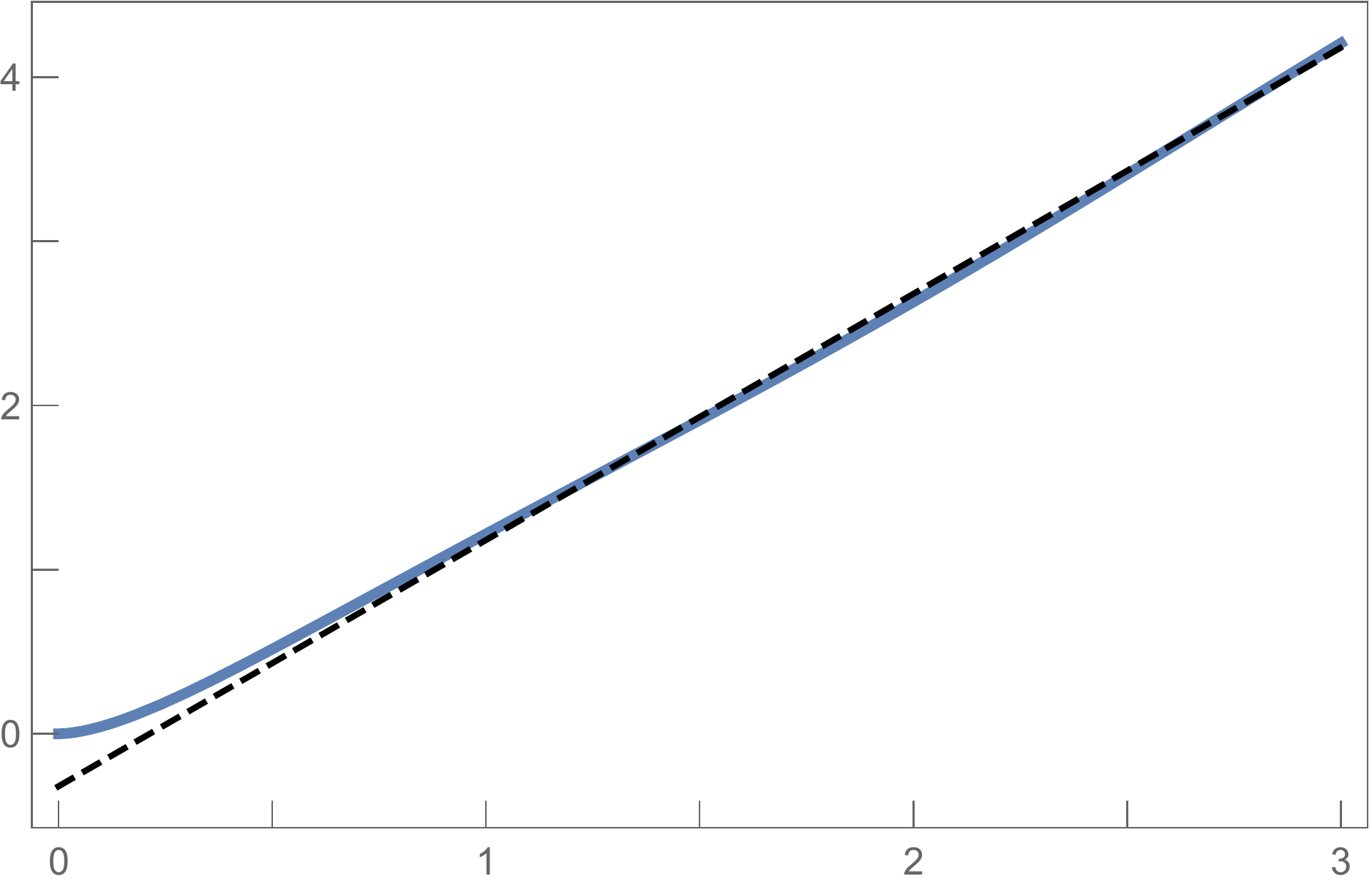}
\hspace{2em}
\includegraphics[height = 12em]{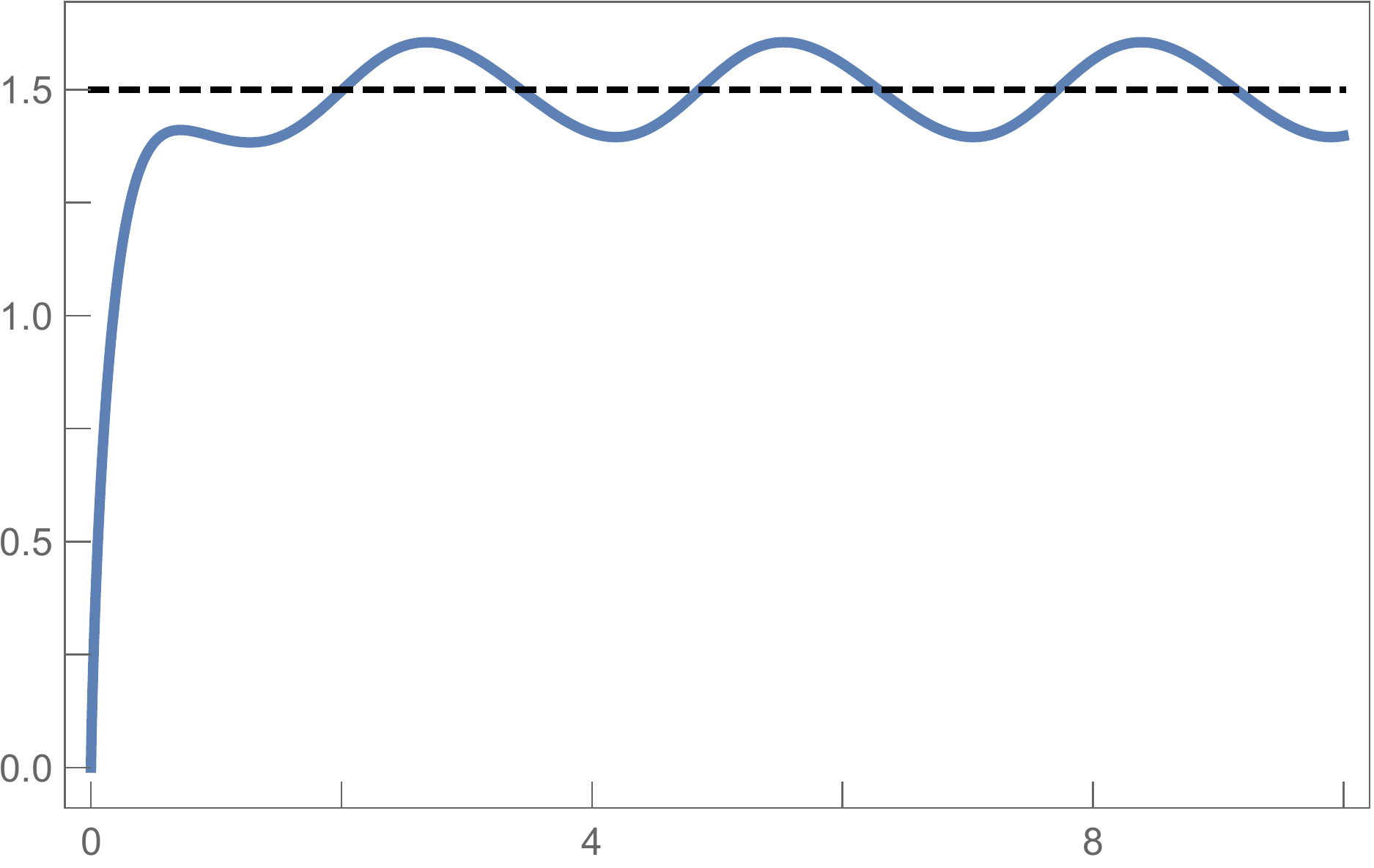}
\caption{Entanglement entropy in the presence of instabilities. Left: entanglement entropy $S_A(|J_t\rangle)$ as a function of time computed using Eq.~(\ref{eq:main}). Right: time derivative of the entanglement entropy. The dashed line is the predicted asymptotic behavior from Eq.~(\ref{eq:dS-average}). The system considered consists of two degrees of freedom. The potential is $\sum_{ij}V_{ij}\phi_i\phi_j=-\frac{1}{2}\lambda^2(\phi_1-\phi_2)^2+\frac{1}{2}\omega^2(\phi_1+\phi_2)^2$. We choose $\lambda\simeq 1.5$ for the unstable direction and $\omega=1.1$ for the stable direction. The subsystem is $A=(\phi_1,\pi_1)$ and the initial state $|J_0\rangle$ is chosen to be the ground state of the potential $V_{ij}=\omega_0^2 \, \delta_{ij}$ with frequency $\omega_0=1.0$. The long-time behavior of the entanglement entropy is linear in time, with rate proportional to $\lambda$ and modulated with frequency $\omega$ around the average. The asymptotic behavior is largely independent of the parameter $\omega_0$ of the initial state and provides numerical support for the relation to the Kolmogorov-Sinai rate $h_{KS}=\lambda$ of Eq.~(\ref{eq:average-lessKS}).}
\label{fig:instab}
\end{figure}

The subleading terms in the long time asymptotics depend on the oscillatory directions with normal frequencies $\omega_k$ and  result in the addition of a second term in Eq.~(\ref{eq:SAt}). This term is bounded and oscillatory so that, defining the time-average $\overline{S}$ over the shortest period $2\pi/\omega_k$, we can write the asymptotics of the rate of entanglement entropy production as 
\begin{equation}
\lim_{t\to \infty}\overline{\frac{\partial S_A(|J_t\rangle)}{\partial t}}\;=\,\sum_{k=1}^{N_*}\delta^A_k\,\delta^B_k\,\lambda_k\,.
\label{eq:dS-average}
\end{equation}
Fig.~\ref{fig:instab} provides a concrete example of this behavior. Note that the classical system described by the Hamiltonian with potential Eq.~(\ref{eq:Vinstab}) has local Lyapunov exponents given by $\pm \lambda_k$ and $\pm \text{i}\,\omega_k$ \cite{zaslavsky2008hamiltonian}. The Kolmogorov-Sinai rate of the system is the sum over the real positive Lyapunov exponents  \cite{pesin1977characteristic},
\begin{equation}
h_{KS}\equiv \sum_{k=1}^{N_I}\lambda_k\,,
\label{eq:}
\end{equation}
a quantity that is in general expected to measure the rate of growth of the coarse-grained entropy of the classical dynamical system \cite{zaslavsky2008hamiltonian,pesin1977characteristic,latora1999kolmogorov,falcioni2005production}. Formula (\ref{eq:SAt}) shows that the long-time behavior of the entanglement entropy in the presence of instabilities is bounded from above by the Kolmogorov-Sinai entropy rate,
\begin{equation}
\lim_{t\to \infty}\overline{\frac{\partial S_A(|J_t\rangle)}{\partial t}}\leq h_{KS}\,.
\label{eq:average-lessKS}
\end{equation}
The equality between the entanglement-entropy rate and $h_{KS}$ is attained when each one of the unstable directions of the potential affects both the subsystem $A$ and its complement $B$, and the number of unstable directions does not exceed the size of the bi-partition, $N_I\leq \min(2N_A,2N_B)$.

\section{Discussion}
We have shown that, treating squeezed vacua as coherent states for the symplectic group, a new formula for their entanglement entropy can be derived, Eq.~(\ref{eq:main}). This formula involves only the complex structure $J$ entering the definition of the squeezed vacuum and its projection $P_A\, J P_A$ to the phase space of the subsystem. We applied this formula to a scalar field on a generic lattice, investigated the relation between the area law and the speed of sound Eq.~(\ref{eq:sound}), and determined a precise relation between the entanglement entropy production in the presence of instabilities and the local Lyapunov exponents of the system, Eq.~(\ref{eq:dS-average}). 

The extension to a continuum quantum field theory of our results is non-trivial because of the presence of infinitely many degrees of freedom. The formulation in terms of complex structures is however well-suited for this analysis: the Fock spaces built on the squeezed vacua $|J_1\rangle$ and $|J_2\rangle$ are unitarily equivalent if and only if $\text{tr}\,(J_1-J_2)^2<\infty$ \cite{berezin:1966,shale1962linear}. In this case Eq.~(\ref{eq:main}) is likely to result in a finite difference of entanglement entropies between the squeezed vacua $|J_1\rangle$ and $|J_2\rangle$, therefore providing a generalization of the results derived in \cite{Holzhey:1994we,Bianchi:2014qua,Bianchi:2014bma}.  Moreover, formula (\ref{eq:main}) goes beyond the notion of geometric entropy \cite{sorkin1983entropy,bombelli1986quantum,srednicki1993entropy}: the subsystem $A$ does not have to be localized in space, it can be for instance a slice in momentum space \cite{Balasubramanian:2011wt}. In particular this remark applies to the analysis of the growth of entanglement entropy in the presence of instabilities of Sec.~\ref{sec:instability} and is directly relevant for the study of entanglement entropy production in inflationary cosmology \cite{Grishchuk:1990bj,Albrecht:1992kf,Polarski:1995jg}.

\acknowledgments
We thank Abhay Ashtekar, Matteo Smerlak and Wolfgang Wieland for numerous discussions on squeezing in quantum field theory and Chris Coleman-Smith for discussions on the Kolmogorov-Sinai entropy. The work of EB is supported by the NSF grant PHY-1404204. NY acknowledges support from CNPq, Brazil.

\appendix

\section{Symplectic algebra and quadratic generators}\label{App:algebra}
$Sp(2N,\mathbb{R})$ is a connected Lie group. In this appendix we describe the algebra of its generators \cite{arnold-givental:1990}. Writing $M=\exp \varepsilon X$ and expanding  $M^a{}_cM^b{}_d\,\Omega^{cd}=\Omega^{ab}$ to the first order in $\varepsilon$ we find the relation $X^a{}_c\,\Omega^{cb}+X^b{}_c\,\Omega^{ac}=0$. This equation can be solved by posing $X=K \Omega$ with $K=(K_{ab})$ a symmetric matrix. Therefore we have that $M=\exp (K \Omega)$ is a symplectic matrix and the symplectic algebra is 
\begin{equation}
\mathfrak{sp}(2N,\mathbb{R})=\{K\Omega| \,K\in\text{Sym}(2N,\mathbb{R})\}\,.
\label{eq:}
\end{equation}
The commutator of two generators is
\begin{equation}
[ K_1 \Omega, K_2 \Omega]=K_3 \Omega
\label{eq:}
\end{equation}
where  $K_3=K_1 \Omega K_2-K_2 \Omega K_1$ is also symmetric. The dimension of the symplectic algebra is
\begin{equation}
\text{dim}\, \mathfrak{sp}(2N,\mathbb{R})=\,\text{dim}\,\text{\emph{Sym}}(2N,\mathbb{R})=N(2N+1)\,.
\label{eq:}
\end{equation}
The generators of the orthogonal group are skew-symmetric and therefore the matrices $K_{\text{is}}$ generating the $\text{\emph{Isot}}(J_0)$ subgroup have the form 
\begin{equation}
K_{\text{is}} = K-J_0K J_0
\label{eq:Kanti}
\end{equation}
for some symmetric matrix $K$. This equation can be solved in block form
\begin{equation}
K_{\text{is}}=
\left(
\begin{array}{cc}
A_1&A_2\\[4pt]
\!-A_2\;\;& A_1
\end{array}
\right)\label{eq:Kis}
\end{equation}
with $A_1\in \text{\emph{Sym}}(N,\mathbb{R})$ and $A_2\in \text{\emph{Skew}}(N,\mathbb{R})$. The antisymmetric matrix $K_{\text{is}}J_0$ generates the $O(2N,\mathbb{R})$ subgroup of the linear symplectic group. As a generic transformation in $\text{\emph{Isot}}(J_0)$ can be written as $R=\exp {K_{\text{is}}J_0}$, we have
\begin{equation}
\text{dim}\, \text{\emph{Isot}}(J_0)=\,\text{dim}\,\text{\emph{Sym}}(N,\mathbb{R})+\,\text{dim}\,\text{\emph{Skew}}(N,\mathbb{R})=N^2\,.
\label{eq:}
\end{equation}

The symplectic matrices in $\text{\emph{Squeeze}}(J_0)$ are symmetric and positive, therefore the generators have the form $K_{\text{sq}} J_0$ with
\begin{equation}
K_{\text{sq}} = K+J_0K J_0
\label{eq:Ksym}
\end{equation}
for some symmetric matrix $K$. This equation can be solved in block form
\begin{equation}
K_{\text{sq}}=
\left(
\begin{array}{cc}
B_1&\;B_2\\
B_2& -B_1
\end{array}
\right)\label{eq:Ksq}
\end{equation}
with $B_1,\,B_2\in \text{\emph{Sym}}(N,\mathbb{R})$. A generic transformation in $\text{\emph{Squeeze}}(J_0)$ can be written as $T=\exp {K_{\text{sq}}J_0}$. Therefore we have
\begin{equation}
\text{dim}\, \text{\emph{Squeeze}}(J_0)=\,2\;\text{dim}\,\text{\emph{Sym}}(N,\mathbb{R})=N(N+1)\,.
\label{eq:}
\end{equation}

The exponential map from the Lie algebra to the group is not surjective. However, any element of the group can be generated by the group multiplication of two elements. Using the polar decomposition $M=T\,R$ 
we can write any symplectic matrix $M$ in a unique way as the product of a matrix in $\text{\emph{Squeeze}}(J_0)$ and a matrix in $\text{\emph{Isot}}(J_0)$,
\begin{equation}
M= \exp({K_{\text{sq}}\, \Omega}) \,\exp({K_{\text{is}}\, \Omega})\,.
\label{eq:}
\end{equation}
This form will be especially important in the following sections where we discuss a unitary representation of the linear symplectic group.\\

In terms of Poisson brackets, Sec.~(\ref{sec:Poisson}), symplectic matrices correspond to linear canonical transformations generated by a quadratic Hamiltonian $H=\frac{1}{2}K_{ab}\xi^a\xi^b$. In particular transformations in $\text{\emph{Isot}}(J_0)$ have generator
\begin{equation}
\frac{1}{2}K_{\text{is}\,ab}\,\xi^a\xi^b\,=\,\frac{1}{2}\Big(A_{1}^{ij}\, p_i p_j \,+A_{1}^{ij}\,q_i q_j\,+A_2^{ij}\,(q_i\,p_j-p_i\,q_j)\Big)
\label{eq:}
\end{equation}
with $A_1\in \text{\emph{Sym}}(N,\mathbb{R})$ and $A_2\in \text{\emph{Skew}}(N,\mathbb{R})$ as in Eq.~(\ref{eq:Kis}). Transformations in $\text{\emph{Squeeze}}(J_0)$ have generator
\begin{equation}
\frac{1}{2}K_{\text{sq}\,ab}\,\xi^a\xi^b\,=\,\frac{1}{2}\Big(B_{1}^{ij}\, p_i p_j \,-B_{1}^{ij}\,q_i q_j\,+B_2^{ij}\,(q_i\,p_j+p_i\,q_j)\Big)
\label{eq:}
\end{equation}
with $B_1, B_2\in \text{\emph{Sym}}(N,\mathbb{R})$ as in Eq.~(\ref{eq:Ksq}).

The analogous expressions in terms of creation and annihilation operators defined in Sec.~(\ref{sec:annihilation}-\ref{sec:EF}) are
\begin{equation}
\frac{1}{2}K_{\text{is}\,ab}\,\xi^a\xi^b=\; \alpha^{ij}\; E_{ij}\,,\qquad\quad
\frac{1}{2}K_{\text{sq}\,ab}\,\xi^a\xi^b=\;\text{i}\,\Big(\frac{1}{2}\beta^{ij}\,F^\dagger_{ij}\;-\frac{1}{2}\beta^{*\,ij}\,F_{ij}\Big)\,,
\label{eq:K-EF}
\end{equation}
where $\alpha\in \text{\emph{Mat}}(N,\mathbb{C})$ is the hermitian matrix $\alpha=A_1+\text{i}A_2$ and $\beta\in\text{\emph{Mat}}(N,\mathbb{C})$ the symmetric matrix $\beta=B_2-\text{i}B_1$. The Bogolyubov coefficients $\Phi_{ij}$ and $\Psi_{ij}$ can be expressed in terms of $\alpha$ and $\beta$ as explained in Sec.~(\ref{sec:EF}).

%In terms of $N\times N$ blocks, a symplectic matrix $M$ and its inverse $M^{-1}=-J_0 M^t J_0$ are given by 
%\begin{equation*}
%M= \Bigg(
%\begin{array}{cc}
%a & b\\
%c & d
%\end{array}
%\Bigg)\,,
%\qquad 
%M^{-1}= \Bigg(
%\begin{array}{cc}
%d^t & -b^t\\
%-c^t & a^t
%\end{array}
%\Bigg)\,.
%\end{equation*}
%with $a, b, c, d\in \text{Mat}(N,\mathbb{R})$  satisfying the conditions
%\begin{equation*}
%a^t d-c^t b=\mathds{1}\,,\qquad a^t c= c^t a\,,\qquad d^t b=b^t d\,.
%\label{eq:}
%\end{equation*}
%The relation between the real matrices $a, b, c, d$  and the Bogoliubov coefficients is
%\begin{equation*}
%a=\text{Re}(\Phi+\Psi)\,,\qquad b=-\text{Im}(\Phi-\Psi)\,,\qquad c=\text{Im}(\Phi+\Psi)\,,\qquad d=\text{Re}(\Phi-\Psi)\,.
%\end{equation*}
%The Bogoliubov coefficients satisfy the relations
%\begin{equation}
%\Phi \Phi^\dagger-\Psi\Psi^\dagger=\mathds{1}\,\qquad \text{and}\qquad \Phi \Psi^t=\Psi\Phi^t\,.
%\label{eq:}
%\end{equation}
%We can move between the real representation $M$ and the complex representation $\check{M}$ by using the relation $M=W^t\check{M}W$. 

\providecommand{\href}[2]{#2}\begingroup\raggedright\endgroup

%\bibliographystyle{JHEP}
%\bibliography{bib-entanglement}

\end{document}